\pdfoutput=1
\documentclass[11pt,twoside,a4paper,cmspaper,final,collab]{cms-tdr}
\def\svnVersion{d044f43}\def\svnDate{2025/02/25}\def\cmsCernNoTag{CERN-EP-2024-214}\def\cmsCernDate{\today}\def\cmsMessage{Published in the Journal of High Energy Physics as \href{http://dx.doi.org/10.1007/JHEP02(2025)089}{\doi{10.1007/JHEP02(2025)089}.}}
\begin{document}\cmsNoteHeader{B2G-23-008}

\newlength\cmsFigWidth
\ifthenelse{\boolean{cms@external}}{\setlength\cmsFigWidth{0.45\textwidth}}{\setlength\cmsFigWidth{0.49\textwidth}}
\newlength\cmsTabSkip\setlength{\cmsTabSkip}{2ex}

\newcommand{\cmsTable}[1]{\resizebox{\textwidth}{!}{#1}}

\providecommand{\PWst}{{\HepParticle{\PW}{}{\ast}}\xspace}
\providecommand{\Pellp}{{\HepParticle{\Pell}{}{+}}\xspace}
\providecommand{\Pellm}{{\HepParticle{\Pell}{}{-}}\xspace}
\newcommand{\pp}{\ensuremath{\Pp\Pp}\xspace}
\newcommand{\Zprime}{\PZpr}
\newcommand{\Zprimexsec}{\ensuremath{\sigma\left(\pp\to\Zprime\right)}\xspace}
\newcommand{\ZprimeBR}{\ensuremath{\mathcal{B}\left(\Zprime\to\PZ\PH\right)}\xspace}
\newcommand{\mZprime}{\ensuremath{m_{\Zprime}}\xspace}
\newcommand{\mZprimeRec}{\ensuremath{m_{\Zprime}^{\text{rec}}}\xspace}
\newcommand{\mTZprime}{\ensuremath{m_{\Zprime}^{\mathrm{T}}}\xspace}
\newcommand{\twobcat}{\ensuremath{2\PQb\,\text{category}}\xspace}
\newcommand{\onebcat}{\ensuremath{{\leq}1\PQb\,\text{category}}\xspace}
\newcommand{\ZH}{\ensuremath{\PZ\PH}\xspace}

\newcommand{\Vjet}{\text{\PV{}+jets}\xspace}
\newcommand{\Wjet}{\text{\PW{}+jets}\xspace}
\newcommand{\Zjet}{\text{\PZ{}+jets}\xspace}

\newcommand{\cF}{\ensuremath{c_{\mathrm{F}}}\xspace}
\newcommand{\gF}{\ensuremath{g_{\mathrm{F}}}\xspace}
\newcommand{\gV}{\ensuremath{g_{\PV}}\xspace}
\newcommand{\cH}{\ensuremath{c_{\PH}}\xspace}
\newcommand{\gH}{\ensuremath{g_{\PH}}\xspace}

\newcommand{\xbb}{\ensuremath{\PX\to\cPqb\cPaqb}\xspace}
\newcommand{\xcc}{\ensuremath{\PX\to\cPqc\cPaqc}\xspace}
\newcommand{\xqq}{\ensuremath{\PX\to\cPq\overline{\cPq}}\xspace}
\newcommand{\HvsQCD}{\ensuremath{\text{HvsQCD}}\xspace}

\cmsNoteHeader{B2G-23-008}
\title{Search for a heavy resonance decaying into a \texorpdfstring{\PZ}{Z} and a Higgs boson in events with an energetic jet and two electrons, two muons, or missing transverse momentum in proton-proton collisions at \texorpdfstring{$\sqrt{s}=13\TeV$}{sqrt(s)=13 TeV}}

\author*[cern]{The CMS Collaboration}

\date{\today}

\abstract{
A search is presented for a heavy resonance decaying into a \PZ boson and a Higgs (\PH) boson. The analysis is based on data from proton-proton collisions at a centre-of-mass energy of 13\TeV corresponding to an integrated luminosity of 138\fbinv, recorded with the CMS experiment in the years 2016--2018. Resonance masses between 1.4 and 5\TeV are considered, resulting in large transverse momenta of the \PZ and \PH bosons. Final states that result from \PZ boson decays to pairs of electrons, muons, or neutrinos are considered. The \PH boson is reconstructed as a single large-radius jet, recoiling against the \PZ boson. Machine-learning flavour-tagging techniques are employed to identify decays of a Lorentz-boosted \PH boson into pairs of charm or bottom quarks, or into four quarks via the intermediate $\PH\to\PW\PWst$ and $\PZ\PZ^{*}$ decays. The analysis targets \PH boson decays that were not generally included in previous searches using the $\PH\to\bbbar$ channel. Compared with previous analyses, the sensitivity for high resonance masses is improved significantly in the channel where at most one \PQb quark is tagged.
}

\hypersetup{
pdfauthor={CMS Collaboration},
pdftitle={Search for a heavy resonance decaying into ZH in events with an energetic jet and two electrons, two muons, or missing transverse momentum},
pdfsubject={CMS},
pdfkeywords={BSM search, ZH final state, jet substructure}}

\maketitle

\section{Introduction} \label{chap:Intro}

The standard model (SM) of particle physics provides a remarkably accurate description of physical 
interactions up to the highest energy scales observed.
Nevertheless, there are indications that extensions to the SM are necessary.
A large variety of theories beyond the SM (BSM) have been proposed and are being tested
by searches at the CERN LHC~\cite{Rappoccio:2018qxp, ATLAS:2024fdw, CMS:2024zqs}.

From an experimental perspective, direct searches for new elementary particles are
typically not sensitive to all the parameters of the underlying theory, 
but instead only to those that affect the production and decay rates of the new particles.
As a consequence, it is sufficient to simplify the complete theory as an effective interaction 
that captures the dynamics of the BSM signal.
One such simplified model is the heavy vector triplet (HVT) model~\cite{Pappadopulo:2014qza},
which describes the production and decay of electroweak spin-1 resonances that arise 
from different theories such as 
weakly coupled extended gauge sectors~\cite{Barger:1980ix, Altarelli:1989ff, Salvioni:2009mt, Grojean:2011vu} and
little Higgs models~\cite{Schmaltz:2005ky,ArkaniHamed:2002qy}, as well as
composite Higgs scenarios~\cite{Contino:2011np,Marzocca:2012zn,Bellazzini:2014yua,Greco:2014aza,Lane:2016kvg}.
Previous searches for a heavy resonance decaying to an SM Higgs (\PH) boson and an electroweak vector boson ($\PV = \PW$, \PZ)
have already been carried out in lepton+jets final states~\cite{CMS:2018ljc, CMS:2021xor,CMS:2021fyk,
CMS:2021klu,CMS:2021itu,ATLAS:2020fry,ATLAS:2022enb}
and in fully hadronic final states~\cite{ATLAS:2024qvg, ATLAS:2020qiz,CMS:2022pjv} by the ATLAS and CMS Collaborations.
The results of these searches are converted into upper limits on the production 
cross sections and lower limits on the heavy-resonance masses in the HVT model~\cite{ATLAS:2024qvg, CMS:2024phk}. 

This article presents a new search for a heavy resonance decaying into a \PZ and an \PH boson,
where the \PH boson decays hadronically and the \PZ boson decays into a pair of
oppositely charged leptons, $\Pep\Pem$ or $\PGmp\PGmm$, or into neutrinos.
The search uses proton-proton (\pp) collision data at a centre-of-mass energy of 13\TeV,
collected by the CMS experiment between 2016 and 2018, corresponding to a total integrated
luminosity of 138\fbinv.
Advanced jet identification algorithms using machine-learning techniques are used
to exploit the \PH boson decays to quarks that have not been generally targeted by previous searches.
Specifically, this search is more sensitive to \PH boson decays into charm quark pairs ($\PH\to\PQc\PAQc$)
and to four-quarks final states ($\PH\to\PV\PV^{*}\to\text{4 quarks}$) than previous searches.
Furthermore, loosely identified \PH boson decays to bottom quark pairs ($\PH\to\bbbar$)
that were not included in previous searches that targeted $\PH\to\bbbar$ decays,
are included in this search.
In a previous search for \ZH resonances by the CMS Collaboration~\cite{CMS:2021fyk},
two event categories based on the \PH boson decay mode were used.
One category focused on $\PH\to\bbbar$ events in which both \PQb quarks were identified experimentally.
The other category targeted the remaining hadronic \PH boson decays.
The combination of these categories resulted in mass exclusion limits of 3.5 and 3.7\TeV,
depending on the benchmark scenario of the HVT model.
The analysis presented in this article demonstrates a significant improvement
in the sensitivity for the category with fewer than two identified \PQb quarks,
while rejecting events with two or more identified \PQb quarks.

The analysis strategy is the following. A pair of oppositely charged leptons or 
the missing transverse momentum are used to identify the $\PZ\to\Pep\Pem$ and $\PGmp\PGmm$,
or $\PGn\PAGn$ decays, respectively.
A large-radius jet recoiling against the measured \PZ boson is used to reconstruct the \PH boson decay.
The internal structure of the \PH boson candidate jet,
resulting from the hadronization of two or four quarks,
and its identified quark flavour content are highly discriminating features employed to separate
a BSM signal from SM background processes~\cite{Larkoski:2017jix, Asquith:2018igt, Kogler:2021kkw}.
Jet substructure and quark flavour identification play an essential role in this analysis, 
and state-of-the-art techniques are employed
to improve the sensitivity for high resonance masses. 
The search is performed in a signal-enriched region by examining the distribution of the 
invariant mass or transverse mass of the reconstructed \ZH system for a 
localized excess over a monotonically decreasing background distribution.
The prediction of the SM background is obtained by fitting a one-dimensional function 
to the observed data and does not rely on simulation. 
The procedure is validated in a background-enriched region with kinematic 
properties similar to those of the signal region.

Tabulated results are provided in the HEPData record for this analysis~\cite{hepdata}.

\section{The CMS detector and object reconstruction}\label{sec:CMS}

The central feature of the CMS apparatus is a superconducting solenoid of 6\unit{m} internal diameter, 
providing a magnetic field of 3.8\unit{T}. Within the solenoid volume are a silicon pixel and strip tracker, 
a lead tungstate crystal electromagnetic calorimeter (ECAL), and a brass and scintillator hadron calorimeter (HCAL), 
each composed of a barrel and two endcap sections. Forward calorimeters extend the pseudorapidity 
coverage provided by the barrel and endcap detectors. Muons are measured in gas-ionization detectors 
embedded in the steel flux-return yoke outside the solenoid. 
More detailed descriptions of the CMS detector, together with a definition of the coordinate 
system used and the relevant kinematic variables, can be found in Refs.~\cite{CMS:2008xjf,CMS:2023gfb}. 

Events of interest are selected using a two-tiered trigger system~\cite{CMS:2016ngn}. The first level, 
composed of custom hardware processors, uses information from the calorimeters and muon detectors
to select events at a rate of around 100\unit{kHz} within a fixed latency of about 4$\mu$s~\cite{CMS:2020cmk}.
The second level, known as the high-level trigger, consists of a farm of processors
running a version of the full event reconstruction software optimized for fast processing,
and reduces the event rate to around 1\unit{kHz} before data storage~\cite{CMS:2016ngn}.

The particle-flow (PF) algorithm~\cite{CMS:2017yfk} aims to reconstruct and identify each particle
in an event, with an optimized combination of information from the various elements of the CMS detector.
The energy of photons is obtained from the ECAL measurement~\cite{CMS:2020uim}. The energy of electrons is determined
from a combination of the electron momentum at the primary interaction vertex as determined
by the tracker, the energy of the corresponding ECAL cluster, and the energy sum of all bremsstrahlung
photons spatially compatible with originating from the electron track~\cite{CMS:2020uim}. 
The energy of muons is obtained from the curvature of the corresponding track~\cite{CMS:2018rym}. 
The energy of charged hadrons is determined
from a combination of their momentum measured in the tracker and the matching
ECAL and HCAL energy deposits, corrected for the response of the calorimeters
to hadronic showers. Finally, the energy of neutral hadrons is obtained from the corresponding
corrected ECAL and HCAL energies.
Jet momentum is determined as the vectorial sum of all particle momenta in the jet,
and is found from simulation to be, on average, within 5 to 10\% of the true momentum
over the entire \pt spectrum and detector acceptance. Additional \pp interactions within
the same or nearby bunch crossings (pileup) can contribute additional tracks and calorimetric energy depositions,
increasing the apparent jet momentum.

The pileup-per-particle identification algorithm (PUPPI)~\cite{Sirunyan:2020foa,Bertolini:2014bba}
is used to mitigate the effect of pileup at the reconstructed PF candidates level, making use of
local shape information, event pileup properties, and tracking information. A local shape variable
is defined, which distinguishes between collinear and soft diffuse distributions of other
particles surrounding the particle under consideration. The former is attributed to particles
originating from the hard scatter and the latter to particles originating from pileup interactions.
Charged particles identified to be originating from pileup vertices are discarded. For each neutral
particle, a local shape variable is computed using the surrounding charged particles compatible with
the primary vertex within the tracker acceptance of $\abs{\eta}<2.5$, where $\eta$ denotes the pseudorapidity.
Both charged and neutral particles are used in the region outside of the tracker coverage. 
The momenta of the neutral particles
are then rescaled according to their probability of originating from the primary interaction vertex
deduced from the local shape variable, superseding the need for jet-based pileup
corrections~\cite{Sirunyan:2020foa}.

The large-radius jets used in this analysis are clustered with the \textsc{FastJet} package~\cite{Cacciari:2011ma}
using the anti-\kt algorithm~\cite{Cacciari:2008gp} with a distance parameter of $R = 0.8$ (AK8 jets). 
The soft-drop algorithm~\cite{Larkoski:2014wba}, 
which is a generalization of the modified mass drop tagger algorithm~\cite{Dasgupta:2013ihk}, 
is used to identify the subjets of a Lorentz-boosted \PH boson candidate jet.
This algorithm, with an angular exponent $\beta = 0$ and a soft-cutoff threshold $z_{\text{cut}}<0.1$,
is applied to AK8 jets reclustered using the Cambridge--Aachen algorithm~\cite{Dokshitzer:1997in,Wobisch:1998wt}, 
and removes soft, wide-angle radiation from the jet. 

Jet energy corrections are derived from simulation studies so that the average measured energy of jets becomes identical
to that of jets at the particle level. In situ measurements of the momentum balance in dijet,
\PGg{}+jet, \PZ{}+jet, and multijet events are used to determine 
any residual differences between the jet energy scale in data and simulation,
and appropriate corrections are made~\cite{CMS:2016lmd}. Additional selection criteria
are applied to each jet to remove jets originating from instrumental effects or
reconstruction failures~\cite{Sirunyan:2020foa}.

The missing transverse momentum vector \ptvecmiss is computed as the negative vector sum of
the transverse momenta of all the PF candidates in an event 
and its magnitude is denoted as \ptmiss~\cite{CMS:2019ctu}. The PUPPI algorithm is applied
to reduce the pileup dependence of the \ptvecmiss observable. The \ptvecmiss is computed
from the PF candidates weighted by their probability to originate from the primary interaction
vertex~\cite{CMS:2019ctu}. The \ptvecmiss is modified to account for corrections to the energy scale
of the reconstructed jets in the event.

\section{Data and simulated samples} \label{sec:DatasetsAndSamples}

Data events were collected with the CMS detector in \pp collisions in the years 2016 to 2018
at $\sqrt{s} = 13\TeV$, corresponding to an integrated luminosity of 138\fbinv.
The data analyzed in this search were recorded using triggers that required
the presence of a single lepton or significant \ptmiss.
To collect events where the \PZ boson decays into a pair of electrons, a combination of
isolated and nonisolated~\cite{CMS:2020uim} electron triggers and photon triggers is used to achieve optimal
efficiency over the whole range of electron energies.
Single-muon triggers without isolation~\cite{CMS:2018rym} criteria are chosen to
avoid efficiency losses in case of very collimated dimuon events. The \pt thresholds for 
the muon triggers ($\pt>50\GeV$) are lower than for electron triggers ($\pt>115\GeV$) without isolation criteria,
such that adding isolated-muon triggers is not necessary in this analysis.

The signal samples are generated using the HVT model~\cite{Pappadopulo:2014qza}.
The couplings of the heavy spin-1 resonance \Zprime to the \PH and \PV bosons ($\gH=\cH \gV$)
and to the fermions ($\gF=g^2 \cF/\gV$) are expressed in terms of
the $\mathrm{SU}(2)_{\mathrm{L}}$ gauge coupling $g$, and of the two dimensionless coefficients
\cH and \cF that control the relative contributions to the total interaction strength \gV.
Universal couplings to different flavours of leptons and quarks are assumed.
Two benchmark scenarios are considered in this analysis~\cite{CMS:2024phk}:
\begin{itemize}
\item Model~A, with $\cH = -0.556$, $\cF = -1.316$, and $\gV = 1$, and
\item Model~B, with $\cH = -0.976$, $\cF =  1.024$, and $\gV = 3$.
\end{itemize}
Model~A has comparable branching fractions into fermions and bosons and 
reproduces a model with a weakly coupled extended gauge theory~\cite{Barger:1980ix}.
Model~B has branching fractions predominantly into bosons and very suppressed fermionic decays.
It mimics a minimal strongly coupled composite Higgs model~\cite{Contino:2011np}.

We generate signal samples for different heavy resonance masses \mZprime 
in the range of 1.4 to 5\TeV, produced via \qqbar annihilation.
The signal simulations use the narrow-width approximation, and we have verified that
the natural widths of the signals are negligible as compared with the detector resolution.
The spin-1 resonance \Zprime decays to a \PZ boson and an \PH boson in all simulated events.
Only \PZ boson decays to charged leptons and neutrinos are simulated,
whereas all possible decays of the \PH boson in the SM are considered.

The main backgrounds in this search originate from \PV boson production 
with additional jets (\Vjet), which includes \Wjet and \Zjet production.
Both signal and \Vjet events are generated with 
\MGvATNLO~2.6.5~\cite{Alwall:2007fs, Alwall:2014hca} at leading order (LO)
in perturbative quantum chromodynamics (QCD).
The \Vjet background is generated with up to four additional partons.
Subdominant background processes include diboson and top quark pair production,
which are generated with \PYTHIA~8.240~\cite{Sjostrand:2014zea} at LO and 
\POWHEG~v2~\cite{Nason:2004rx,Frixione:2007vw,Frixione:2007nw,Alioli:2010xd,Re:2010bp} 
at next-to-LO (NLO), respectively.
The cross section for the \ttbar background is adjusted to a prediction at 
next-to-NLO (NNLO) accuracy in perturbative QCD, 
using a next-to-next-to-leading logarithmic soft-gluon resummation,
obtained with the \textsc{Top++}~2.0 program~\cite{Czakon:2011xx}.

The parton showering and hadronization processes
are simulated with \PYTHIA, the underlying event simulation uses the CP5 tune~\cite{Sirunyan:2019dfx},
and the NNPDF~3.1~\cite{Ball_2017} NNLO parton distribution function (PDF) set is employed.

All simulated samples are processed through a \GEANTfour-based~\cite{GEANT4} simulation of the CMS detector.
To simulate the effect of pileup collisions, additional inelastic events are generated using \PYTHIA 
with a total inelastic cross section of 69.2\unit{mb}~\cite{Sirunyan:2018nqx}
and superimposed on the hard-scattering events.
The generated number of \pp interactions is corrected such that
the simulated distribution in the number of primary vertices matches that observed in the data.

\section{Event selection and reconstruction}
\label{chap:Selections}

The final state targeted in this search consists of an energetic AK8 jet from the hadronic \PH boson decay,
recoiling against a pair of charged leptons or \ptmiss from the \PZ boson decay.
Events in the charged-lepton channels $\Pellp\Pellm$ ($\Pell=\Pe,\PGm$) are placed
into mutually exclusive categories based on the flavour of the reconstructed leptons.
Each category requires the presence of two leptons of the same flavour and opposite sign.
Additionally, events with different-flavour leptons are discarded.
In the $\Pellp\Pellm$ channels, events are required to have at least two leptons with $\pt>52\GeV$ and $\abs{\eta}<2.4$ each.
The invariant mass of the dilepton pair must be within 81--101\GeV,
consistent with the \PZ boson mass.
In events with more than one pair of leptons, corresponding to less than 0.1\% of the cases,
the pair with mass closest to the nominal \PZ boson mass is chosen.
The \PZ boson four-momentum is reconstructed from the dilepton system.
A heavy resonance decay would result in a large \pt for the \PZ boson,
satisfying $\pt(\PZ)>200\GeV$.
The leptons are required to be isolated, and to pass the ``loose'' criteria
of the cutoff-based algorithm and the ``tight'' criteria of the ``track-based'' algorithm
for electrons and muons, respectively. The isolation methods and algorithms
are detailed further in Refs.~\cite{CMS:2020uim, CMS:2018rym}.
When building the isolation variable for one lepton, the other lepton is not considered
in the list of particles inside the isolation cone.
This procedure prevents the rejection of events from very collimated \PZ boson decays
where the two leptons are reconstructed within each other's isolation cone.
In \Zjet events, the two leptons are expected to have a large angular separation compared with signal events,
where the angular separation is defined as
$\Delta R(\ell_1,\ell_2)=\sqrt{\smash[b]{[\Delta\eta(\ell_1,\ell_2)]^2 + [\Delta\phi(\ell_1,\ell_2)]^2}}$,
with $\Delta\eta$ and $\Delta\phi$ denoting the differences in $\eta$ and the azimuthal angle $\phi$.
We require that the angular separation between the two leptons must be less than 0.45.
The neutrino channel requires the absence of any electrons or muons, and $\ptmiss>250\GeV$.
A veto on the presence of additional leptons is imposed on leptons passing the same quality
criteria of each of the charged-lepton channels.

To provide a \PH boson candidate, events must contain at least one AK8 jet with $\pt>200\GeV$ and $\abs{\eta}<2.4$.
The \PH boson candidate is reconstructed as a single AK8 jet, referred to as \PH jet, and is required to
have a large azimuthal separation from the reconstructed \PZ boson system, $\Delta\phi(\PH,\PZ)>2$.

The previous CMS result~\cite{CMS:2021fyk} was obtained by using a combination of two mutually
exclusive event categories, based on the number of \PQb-tagged subjets of a boosted \PH boson candidate jet
that includes all decay products of the \PH boson decay chain.
The two categories were obtained by separating events into cases where the \PH jet
contains two (\twobcat) or one or fewer (\onebcat) \PQb-tagged subjets.
While the first category targets $\PH\to\bbbar$ decays, the latter
category includes all remaining hadronic \PH boson decays,
including decays to \PQb quarks that failed identification by the \textsc{DeepCSV} algorithm~\cite{BTV-16-002}.
The working point of the algorithm was chosen to correspond to a 10\% misidentification rate for jets 
originating from light quarks or gluons, and to yield an efficiency between 80 and 90\% for selecting \PQb-quark-initiated jets.   
This led to an optimal sensitivity of the \twobcat. 
In this analysis, we veto events that contain \PH boson candidates with two \PQb-tagged subjets using the 
identical \textsc{DeepCSV} algorithm with the same working point.
The requirement of ${\leq}1$ \PQb-tagged subjets results in a rejection of
40--70\% of $\PH\to\bbbar$ events, depending on the resonance mass, while retaining 80--90\% of
both $\PH\to\ccbar$ and $\PH\to\qqbar\qqbar$ events.
This requirement ensures that the selected events are statistically independent 
of the \twobcat of the previous CMS analysis~\cite{CMS:2021fyk}. 
We then reanalyze the data in the \onebcat with enhanced techniques to improve the sensitivity.

\begin{figure}[t]
\centering
\includegraphics[width=\cmsFigWidth]{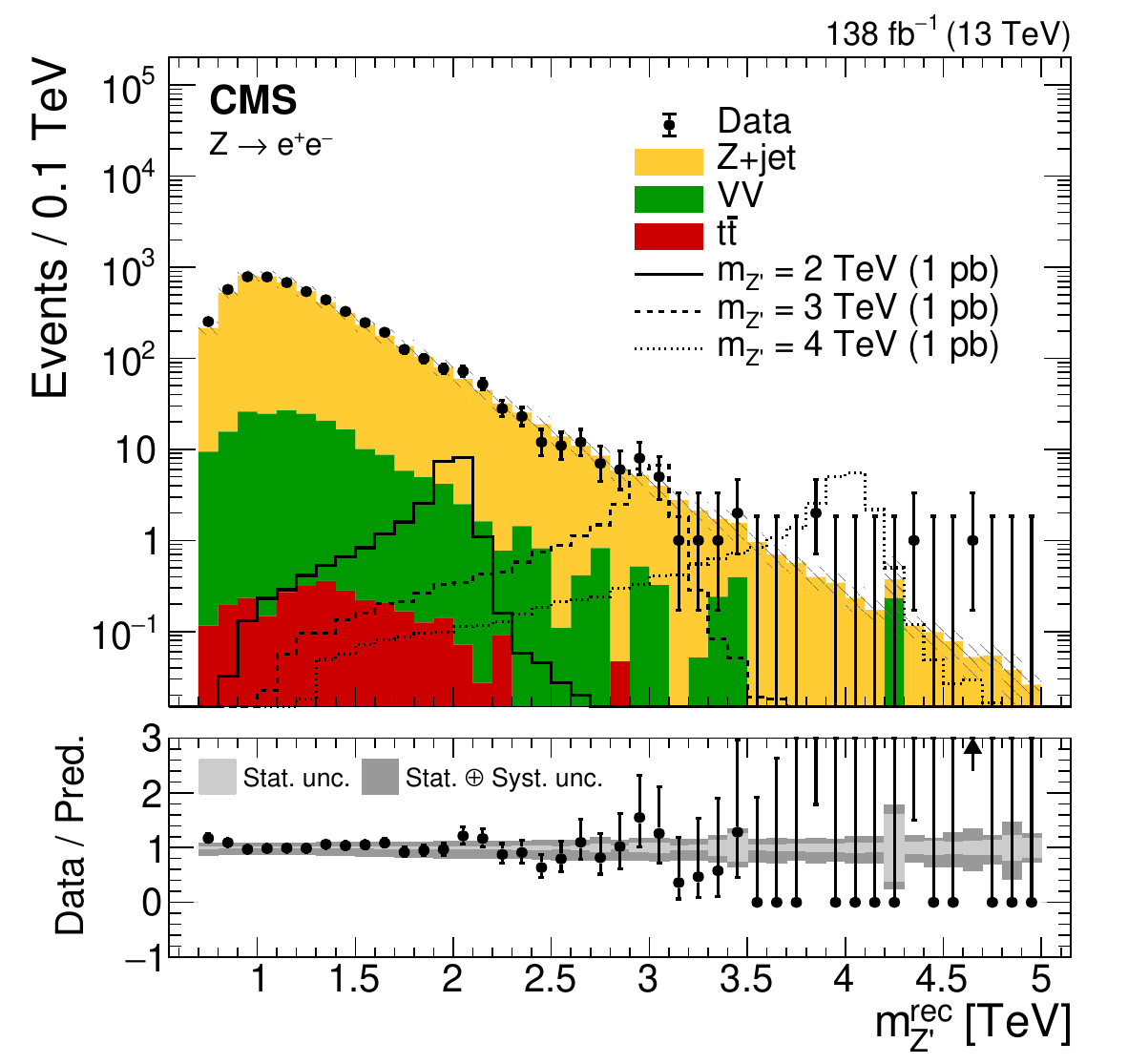}
\includegraphics[width=\cmsFigWidth]{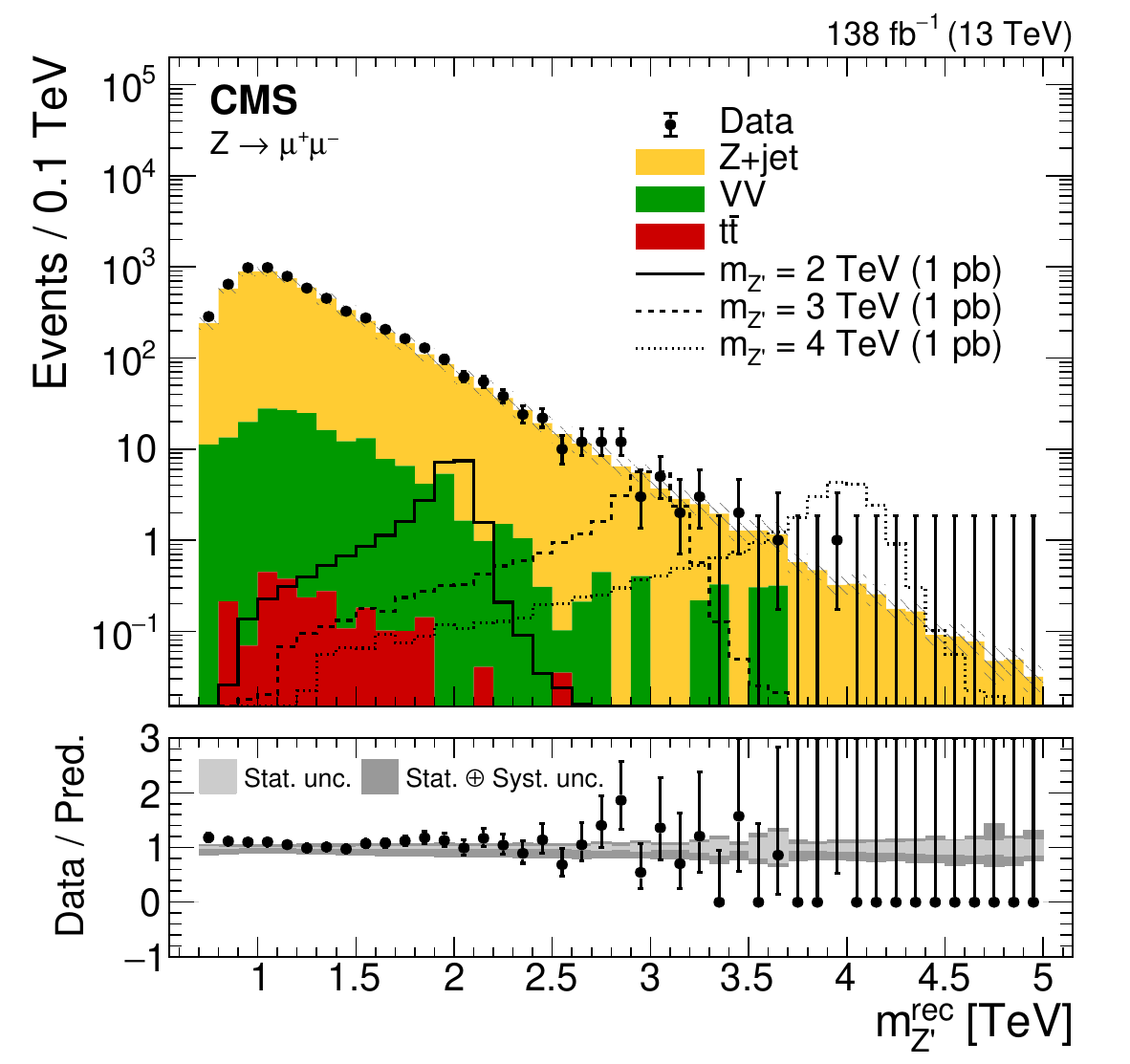}\\
\includegraphics[width=\cmsFigWidth]{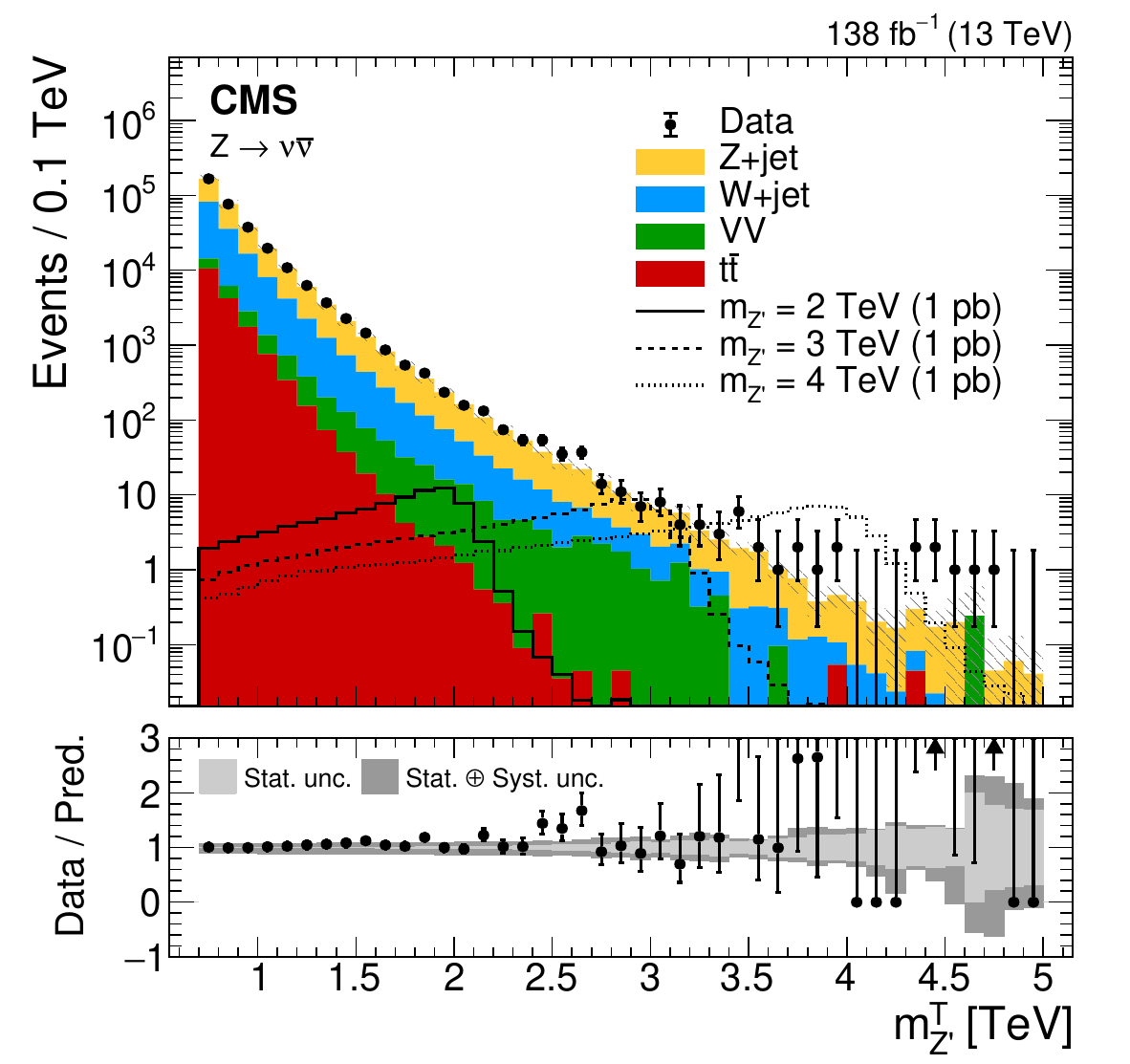}
\caption{Distributions in \mZprimeRec for the dielectron (upper left), dimuon (upper right),
and in \mTZprime for the neutrino (lower) channels, after the kinematic selections.
The data are compared with simulation. The ratios of the data to the total SM background are shown in the lower panels,
where the statistical and total uncertainties are displayed as grey regions.
The signal distributions are shown for an arbitrary cross section of 1\unit{pb}.}
\label{fig:mass_Zprime_control_region}
\end{figure}
The primary source of background arises from \Vjet production.
For such events, the \PV boson is produced in association with a highly energetic jet,
which is misidentified as an \PH boson jet. In the neutrino channel, the second largest source of
background events consists of \Wjet events where the charged lepton from the leptonic \PW
boson decay is not reconstructed.
The signal would appear as a local enhancement in the distribution of the reconstructed invariant mass of the $\PZ\PH$ system,
denoted by \mZprimeRec, in the charged-lepton channels. In the neutrino channel, the sensitive observable is the transverse mass,
\begin{equation}
\mTZprime=\sqrt{2 \pt^{\PH}\ptmiss [1-\cos\Delta\phi(\ptvec^{\PH},\ptvecmiss)]},
\end{equation}
where $\pt^{\PH}$ is the transverse momentum of the \PH jet and
$\Delta\phi(\ptvec^{\PH},\ptvecmiss)$ is the angular difference in azimuth between the \ptvec of the \PH jet and $\ptvecmiss$.
Figure~\ref{fig:mass_Zprime_control_region} shows the distributions in \mZprimeRec
for the charged-lepton channels (upper) and the distribution in \mTZprime for the neutrino channel (lower).
The shapes and normalizations of these distributions are generally well described by the simulation.
In the next step, the signal-to-background ratio is improved by selecting events that
contain a jet consistent with originating from the decay of a boosted \PH boson.

The quark flavour tagging of subjets in the selected \PH jet provides discrimination
between the signal and background processes. In \Vjet events, jets are produced from initial-
or final-state radiation and originate most commonly from the fragmentation of gluons
or light-flavour quarks.
Conversely, jets originating from decays of boosted \PH bosons are characterized
by a more complex substructure and a larger component from the fragmentation of heavy-flavour quarks.

We improve the signal-to-background ratio by applying the
\textsc{ParticleNet-MD}~\cite{Qu:2019gqs,CMS-PAS-BTV-22-001} algorithm to the \PH jet.
This heavy-flavour jet identification algorithm is based on a neural network (NN),
and provides several classification output nodes, corresponding to probability-like scores $p$.
In particular, the mass-decorrelated~\cite{Dolen:2016kst} version of this algorithm is used in this work. 
This version ensures that the background distribution is not altered by 
a selection based on the \textsc{ParticleNet-MD} algorithm score~\cite{CMS:2020poo}.
A combination of these probabilities is used to identify a region enriched
in events with \PH-boson-initiated jets and depleted in events with light-quark- and gluon-initiated jets.
The chosen variable is defined as
\begin{equation}
\HvsQCD=\frac{p(\xbb) + p(\xcc)+ p(\xqq)}{p(\xbb) + p(\xcc)+ p(\xqq) + p(\text{QCD})},
\end{equation}
where \PX is a spin-0 particle with unknown mass decaying to single jets containing pairs of
\PQb quarks, \PQc quarks, and light quarks; and $p(\text{QCD})$ is the probability
that the jet is produced in a QCD multijet event.
Additional details on the \textsc{ParticleNet-MD} network and its calibration
can be found in Refs.~\cite{Qu:2019gqs,CMS-PAS-BTV-22-001}.

The \HvsQCD discriminant is used to define our signal region (SR) by requiring the score
of the AK8 jet to be greater than 0.95, which corresponds to a background misidentification rate of 1\%.
The \xbb probability alone provides substantial background rejection
while selecting a significant fraction of signal events.
The inclusion of the \xcc and \xqq probabilities further improves the overall signal selection
and enables precise calibration of the identification and misidentification rates of the tagger.
This novel technique enables a generalized approach to the tagging of hadronic final states,
remaining largely independent of specific branching fractions,
which allows for a more model-independent search.

Given the dominant branching fraction of $\PH\to\bbbar$,
approximately 60\% of the selected signal events consist of $\PH\to\bbbar$ decays
despite the rejection of jets with two \PQb-tagged subjets.
The selection based on the \HvsQCD discriminant retains most of the remaining $\PH\to\bbbar$
events and provides a substantial background rejection at the same time.
After applying the selection on the \HvsQCD score, the observed relative contributions of
signal events from $\PH\to\bbbar$, $\qqbar\qqbar$, \ccbar, and $\PGt\PGt$ decays are about 
60\%, 20\%, 15\%, and 5\%, respectively.
The analysis selection increases the fraction of signal events from $\PH\to\ccbar$ and $\qqbar\qqbar$ decays
by factors of five and three, respectively, as compared with the SM branching fractions.
Specifically, 25\% of the $\PH\to\qqbar\qqbar$ events have all four quarks matched
to the same AK8 jet, whereas in the remaining 75\%, at most three quarks are matched.
A significant fraction of these quarks matched to the AK8 jet are \PQc quarks, which results in an increased \HvsQCD tagging probability.

\begin{figure}[tb]
\centering
\includegraphics[width=0.6\textwidth]{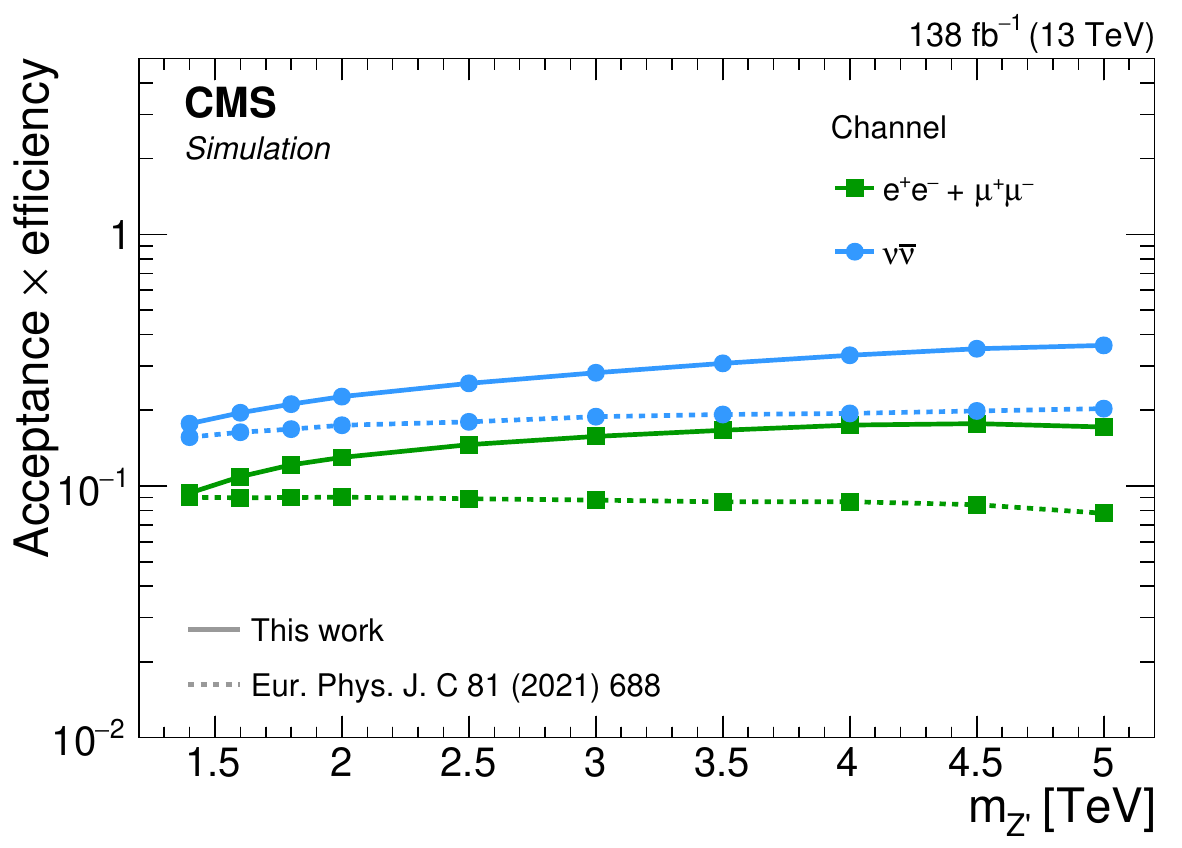}
\caption{The product of signal acceptance and efficiency for signal events as a function of \mZprime
for the charged-lepton and neutrino channels in the SR.
The efficiency is calculated with respect to \PZ boson decays to charged leptons and neutrinos
for the charged-lepton and neutrino channels, respectively.
For comparison, the results from the \onebcat of the previous CMS search in the \ZH
channel~\cite{CMS:2021fyk} are shown as dashed lines.
\label{fig:signal_efficiency}
}
\end{figure}

The product of the geometrical acceptance and the selection efficiency for different channels
in the SR for all simulated signal samples is shown in Fig.~\ref{fig:signal_efficiency}.
In the neutrino channel, the efficiency is calculated with respect to \PZ boson decays to neutrinos,
and in the charged-lepton channels with respect to \PZ boson decays to electrons, muons, and \PGt leptons.
We observe a gain in the product of signal acceptance and efficiency of about a factor of two
as compared with the \onebcat of the previous CMS analysis~\cite{CMS:2021fyk}.
Using the flavour and substructure of the selected \PH jet
allows us to achieve the same level of background suppression with a gain in signal efficiency.

\section{Signal and background modelling}\label{sec:modelling}

The search is performed by examining the distributions in \mZprimeRec and \mTZprime 
in the SRs for localized excesses over monotonically decreasing background distributions.
Each SM background prediction is obtained by fitting the free parameters of a one-dimensional
function to the data.
This procedure reduces statistical fluctuations and systematic uncertainties associated with the 
simulation of background processes.
The parametric background model is tested in a background-enriched validation region (VR), 
defined by inverting the selection on the \HvsQCD discriminant.
Since the kinematic properties of the SRs and VRs are similar, the functional form of the background 
prediction can be tested on data in the VRs before examining the SRs.

\begin{figure}[tp]
\centering
\includegraphics[width=0.44\textwidth]{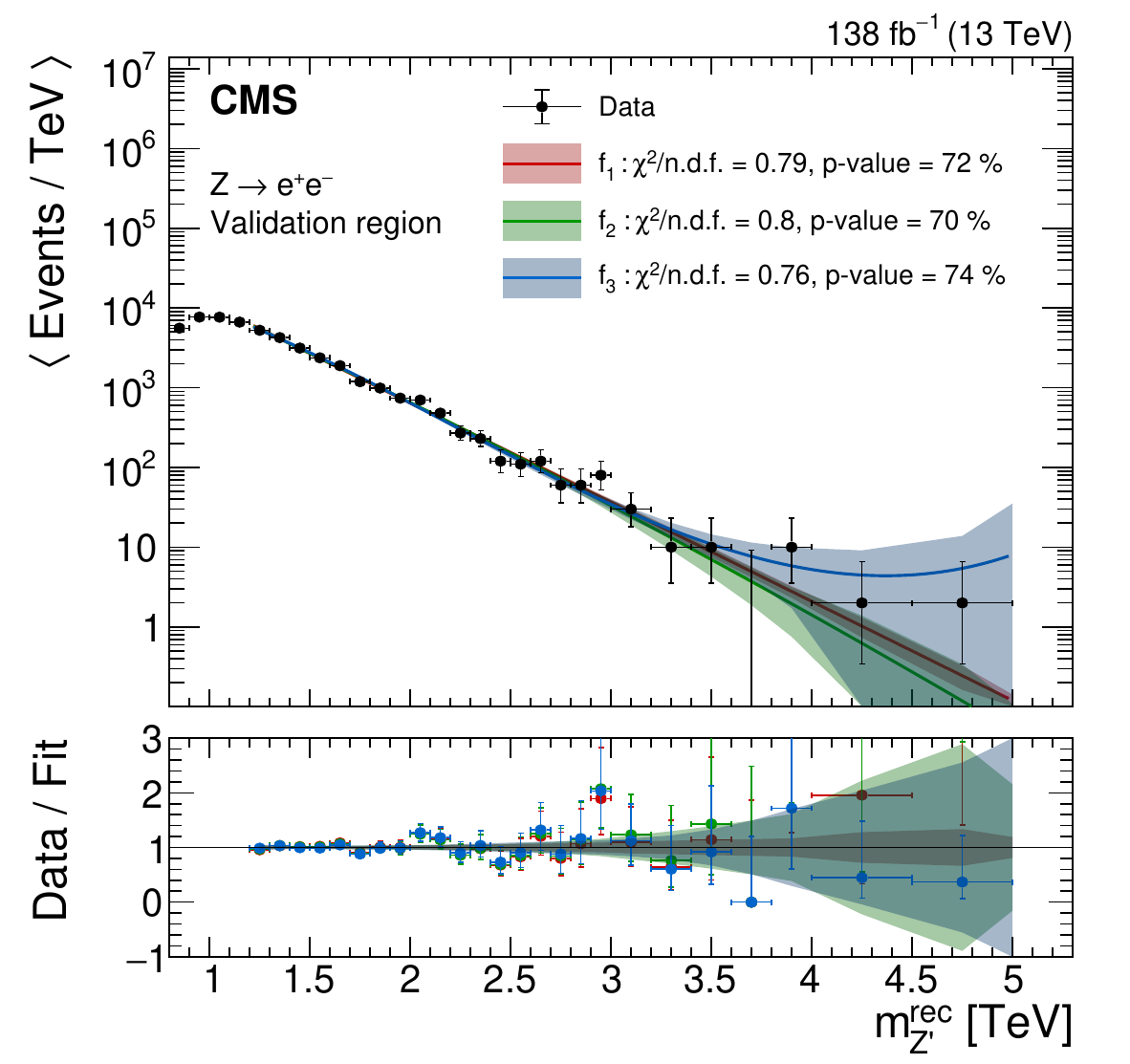}
\includegraphics[width=0.44\textwidth]{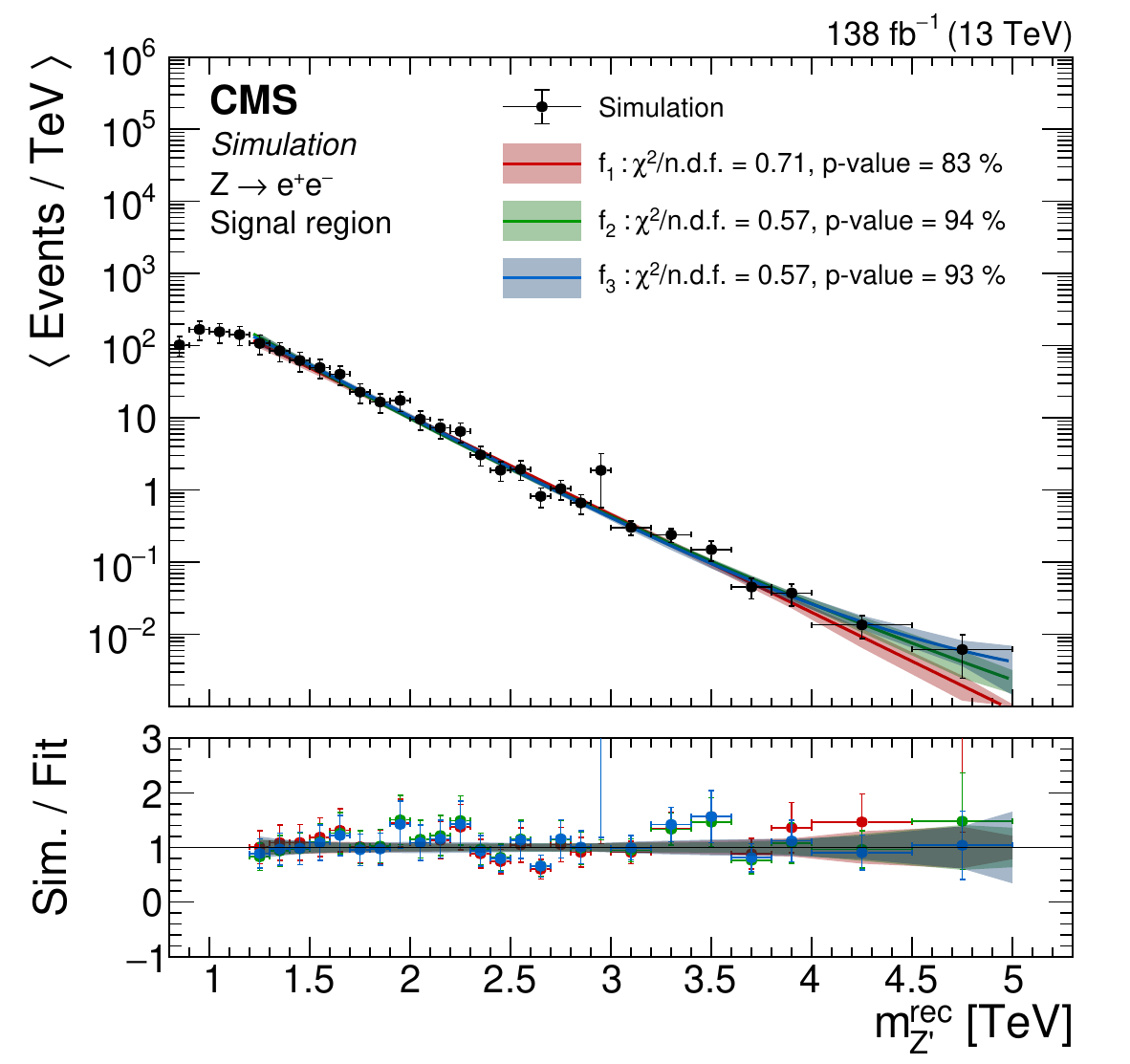}\\
\includegraphics[width=0.44\textwidth]{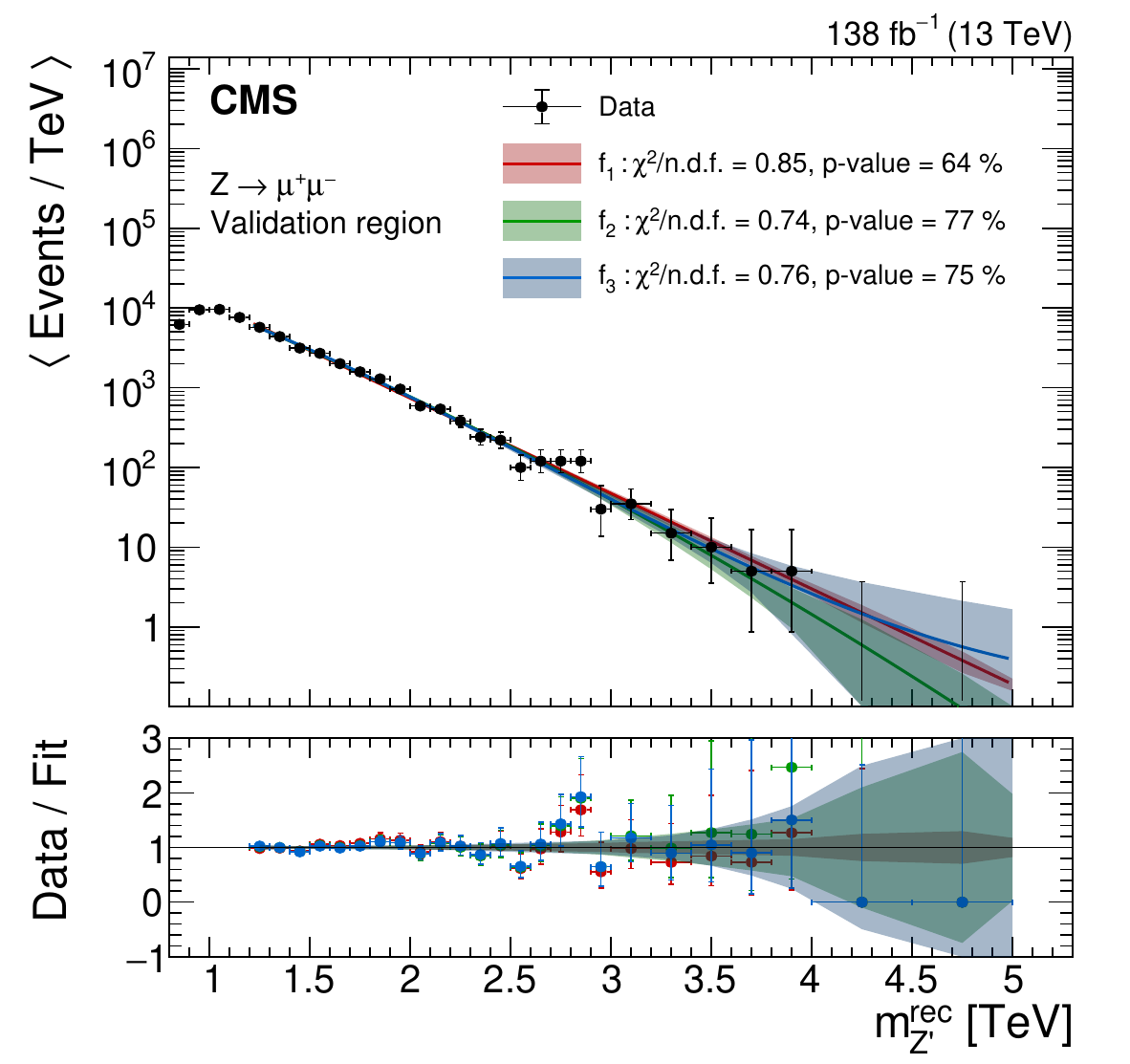}
\includegraphics[width=0.44\textwidth]{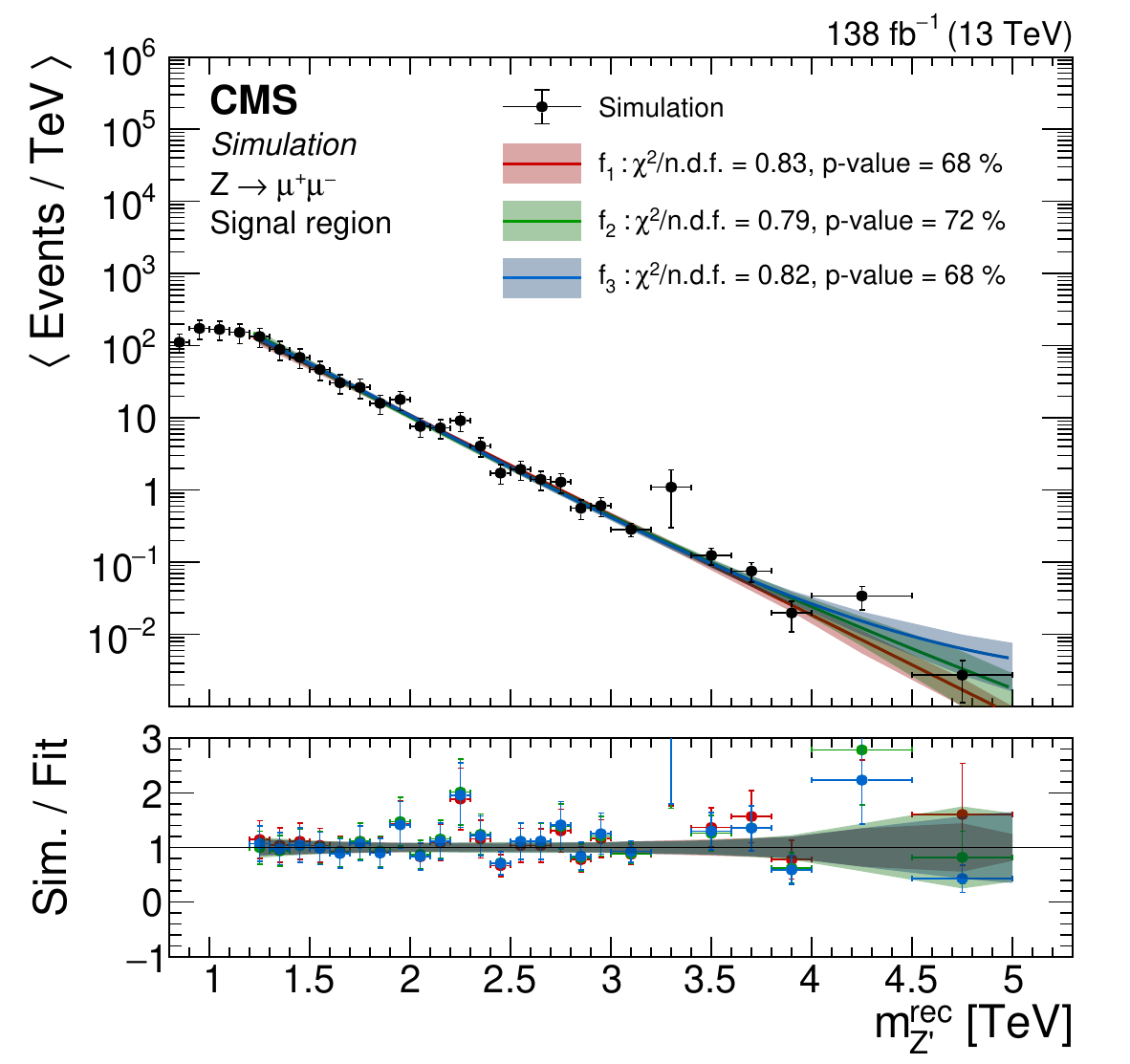}\\
\includegraphics[width=0.44\textwidth]{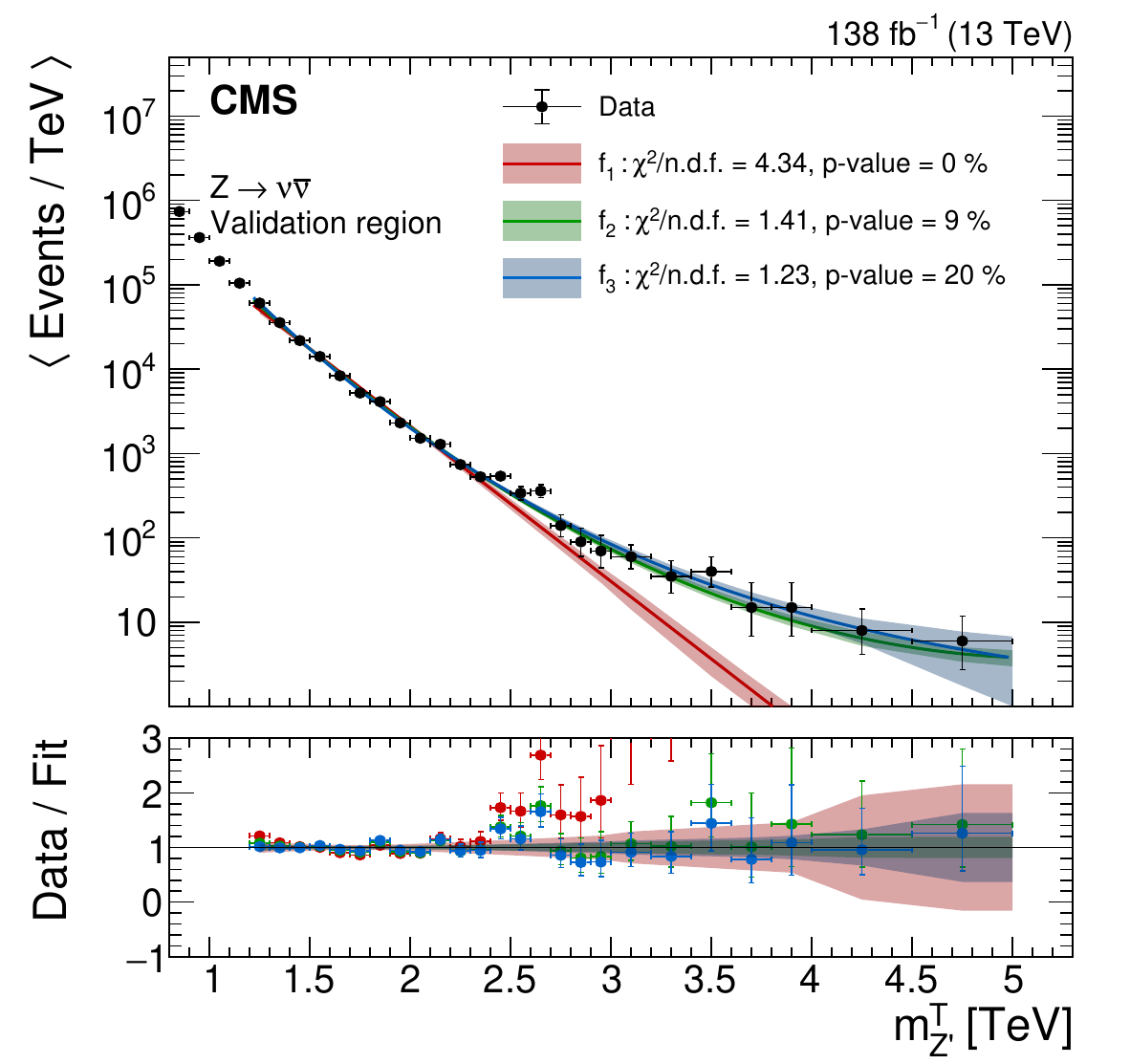}
\includegraphics[width=0.44\textwidth]{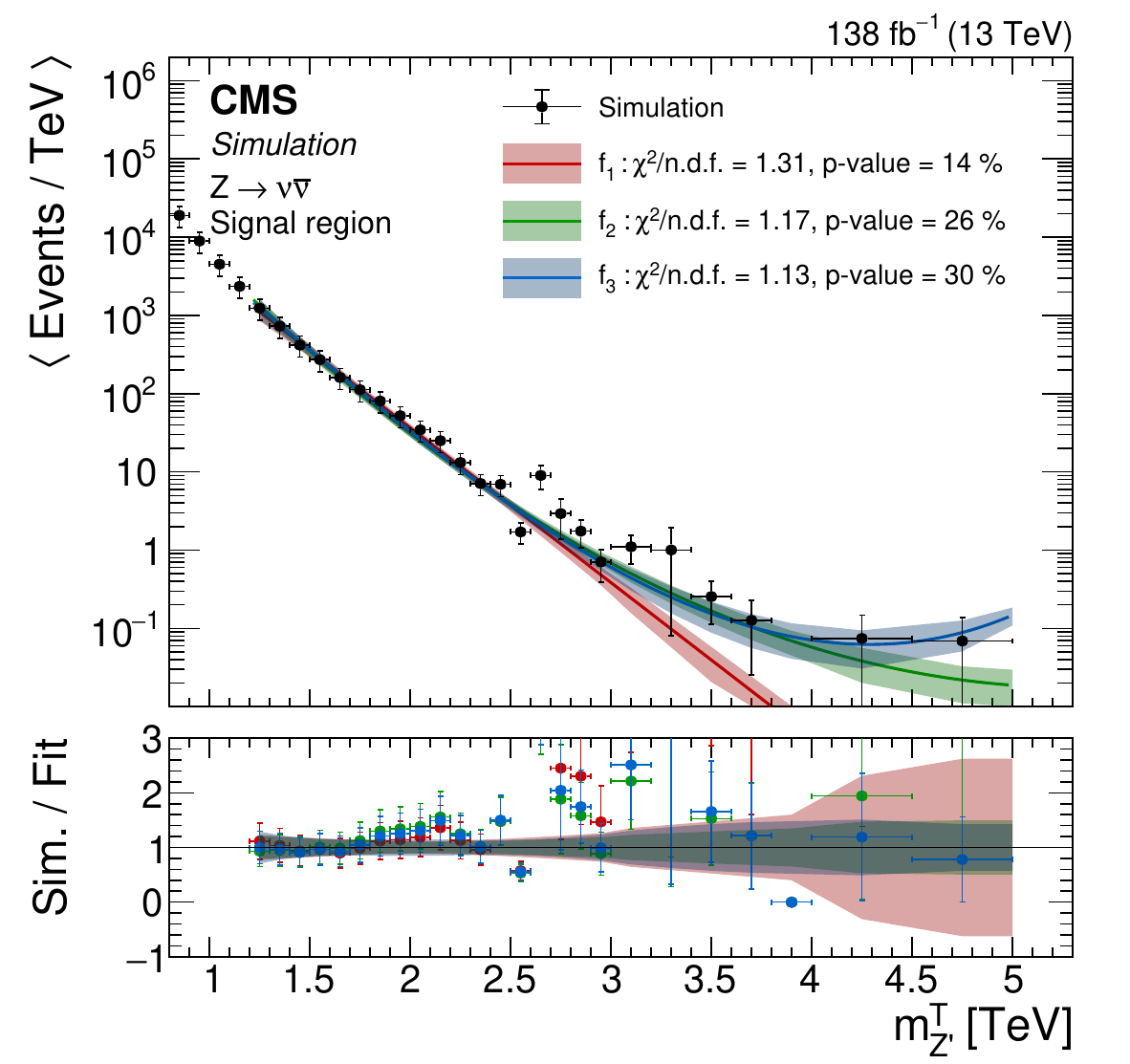}\\
\caption{Fits of the background functions to the \mZprime and \mTZprime distributions in data in the VRs (left)
and simulation in the SRs (right) for the electron (upper), muon (middle), and neutrino (lower) channels.
The number of events in each bin is divided by the bin width.
In the panels below the distributions, the ratios of data to the background functions are displayed.
The ratios to the functions $f_1$, $f_2$, and $f_3$, are shown by red, green, and blue points, respectively.
The shaded colored areas represent the statistical uncertainty from the fit.
The $\chi^{2}$ values per number of degrees of freedom ($\chi^{2}$/n.d.f.) and 
the corresponding $p$-values are provided for each fit.
\label{fig:VR_fits}
}
\end{figure}
The functional form used to fit the background components in the \mZprimeRec and \mTZprime distributions 
is given by
\begin{equation}
f_N(x)=\exp \left( \sum_{i=0}^N p_i x^i \right),
\label{eq:background}
\end{equation}
where $N$ represents the degree of the polynomial function in the exponent and the 
coefficients $p_i$ are free parameters.
The fit range excludes the kinematic turn-on in the \mZprimeRec and \mTZprime distributions, 
resulting from the selection criteria on the leptons, \ptmiss, and AK8 jet. The lower bound of the fit 
range is chosen uniformly across all channels at 1.2\TeV, such that the peak position of 
the signal with the lowest resonance mass considered in this search, $\mZprime = 1.4\TeV$, 
is about three times the mass resolution higher than the lower bound.

An $F$-test~\cite{Ftest} is used to determine the minimum number of free parameters needed to 
describe the distributions of data in the VRs and of the simulations in the SRs and VRs.
The result is $N=2$, where functions with $N=1$ are not able to describe the high-mass tail in the 
VR of the neutrino channel, and $N=3$ does not lead to a statistically significant improvement of the fit quality.
Fits of the background functions to the distributions in \mZprimeRec and \mTZprime are shown in
Fig.~\ref{fig:VR_fits} for simulated events in the SRs and collision data in the VRs.
The simulation in the SRs and the data in the VRs have higher statistical precision than the data in the SRs, 
giving confidence that the chosen background function can reliably describe the background 
shape and normalization in the SRs.
The best fit parameters and their uncertainties obtained from the simulation in the SRs and the data in the VRs
are not used further in the analysis.
The $F$-test is repeated later using the data in the SR, confirming the choice of $N=2$.

The line shape of the simulated signals in \mZprimeRec and \mTZprime can be described by a Gaussian core
centred around the generated \mZprime and a long tail towards smaller masses.
These asymmetric tails result from the off-shell production of the heavy resonance and 
from decay particles not reconstructed in the jet.
The distributions in \mZprimeRec and \mTZprime of the signals are modelled
with a double-sided Crystal Ball function~\cite{Oreglia:1980cs,Gaiser:1982ua,Skwarnicki:1986xj},
which accurately describes the line shape of the simulated signals.

\section{Systematic uncertainties} \label{sec:Systematics}

The analysis sensitivity is dominated by the statistical uncertainty from the number of events in the SRs.
The background parameters from Eq.~\eqref{eq:background} have the largest impact on the results 
of the analysis because the data in the SRs only loosely constrain the background model 
in the high-mass tails of the distributions.
The statistical uncertainties in the background predictions for the charged-lepton SRs range from
15\% for $\mZprime=1.5\TeV$ to 100\% for $\mZprime=3.0\TeV$.
Likewise, for the neutrino SR, these uncertainties vary between 5\% and 45\%.
The uncertainties in the background predictions translate into uncertainties in the signal normalization 
ranging from about 15\% up to about 35\%, depending on \mZprime and the signal cross section.
The systematic uncertainty due to the choice of the background model
is negligible compared with the statistical uncertainty.

\begin{table}[t]
\topcaption{Sources of signal systematic uncertainties considered in this analysis,
and their effect on the signal normalization.
The uncertainty ranges correspond to different signal masses.
These uncertainties are subleading compared with those in the background predictions.
}
\centering
\cmsTable{
\begin{tabular}{lc}
Source & Uncertainty \\
\hline
\textsc{ParticleNet-MD} \PH jet identification  & 2.0--5.0\%\\
\PQb tagging veto                              & 0.4--1.0\%\\
Jet energy scale and resolution                & 0.2--2.0\%\\
Pileup                                         & 0.3--1.8\%\\
Integrated luminosity                          & 1.6\%\\
L1 ECAL trigger inefficiency                   & 0.3--0.8\%\\
\end{tabular}
\hspace*{5pt}
\begin{tabular}{lc}
Source & Uncertainty \\
\hline
Trigger                         & 0.9--1.5\%\\
Muon identification             & 0.1--0.3\%\\
Electron identification         & 5.2--6.0\%\\
Lepton reconstruction           & 0.9--2.0\%\\
PDF                             & 0.3--13.4\%\\
Ren. and fact. scales           & 6.6--17.2\%\\
\end{tabular}
}
\label{tab:systematic_split}
\end{table}
Several additional sources of systematic uncertainty are considered, as these can affect 
the normalization and lineshapes of the signal distributions.
These are briefly described in the following.
Table~\ref{tab:systematic_split} summarizes the effect of each source of systematic uncertainty 
on the signal normalization.

Differences in the selection efficiencies between data and simulation are corrected with
data-to-simulation scale factors (SFs).
The experimental uncertainties are evaluated by varying the SFs up and down by one standard deviation
for each uncertainty source and propagating the resulting change through the analysis.
The resulting systematic uncertainties are treated as fully correlated across the channels.
The largest experimental systematic uncertainty is related to the \PH jet identification 
using the \textsc{ParticleNet-MD} algorithm and amounts to 2--5\%. The uncertainty related to the 
rejection of double-\PQb-tagged jets amounts to 0.4--1.0\%.
The jet energy scale and resolution uncertainties amount to up to 2\% each.
These uncertainties affect all channels equally and therefore have the largest impact on the analysis.
Uncertainties in the trigger efficiency, lepton identification, and reconstruction are considered
in the charged-lepton channels, leading to uncertainties of 5--6\% for the electron channel
and up to 2\% for the muon channel.
Additional systematic uncertainties originate from estimations of the pileup reweighting,
the integrated luminosity~\cite{CMS:LUM-17-003,CMS:LUM-17-004,CMS:LUM-18-002}, 
and L1 ECAL trigger inefficiency due to detector timing issues~\cite{CMS:2020cmk}, each having an effect of less than 2\%.
While most of these uncertainties affect both the normalization and shape of the signal distributions, 
the effect on the normalization is the dominant effect.

In addition, we consider theoretical uncertainties related to the production of the heavy resonance.
The signal cross sections are affected by the choice of the QCD renormalization 
and factorization scales, as well as by the uncertainties in the PDFs used to generate the signal samples.
The effect of these is estimated following the PDF4LHC recommendations~\cite{Butterworth:2015oua}.
We find that acceptance effects from these uncertainties are small as compared with
other uncertainties and we do not consider them as part of the signal uncertainties.
Instead, we include these theoretical uncertainties in the predictions of the production cross sections 
for Models~A and B, based on Ref.~\cite{Pappadopulo:2014qza}.
The size of these is 6.6--17.2\% and 0.3--13.4\% for the scale variations and PDFs, respectively.
The uncertainties increase at higher signal masses.

\section{Results} \label{sec:Results}

The modified frequentist approach~\cite{CMS-NOTE-2011-005,CLS1,CLS2},
known as the \CLs criterion with the profile likelihood ratio as the test statistic,
is used in this search to set 95\% confidence level (\CL) upper limits on the product of production cross section
\Zprimexsec and branching fraction \ZprimeBR.
We use the asymptotic approximation to the profile likelihood test statistic~\cite{Cowan:2010js}.
The following results have been determined using the CMS statistical analysis tool \textsc{Combine}~\cite{CAT-23-001},
which is based on the \textsc{RooFit}~\cite{Verkerke:2003ir} and \textsc{RooStats}~\cite{Moneta:2010pm} frameworks.

\begin{figure}[tp!]
\centering
\includegraphics[width=\cmsFigWidth]{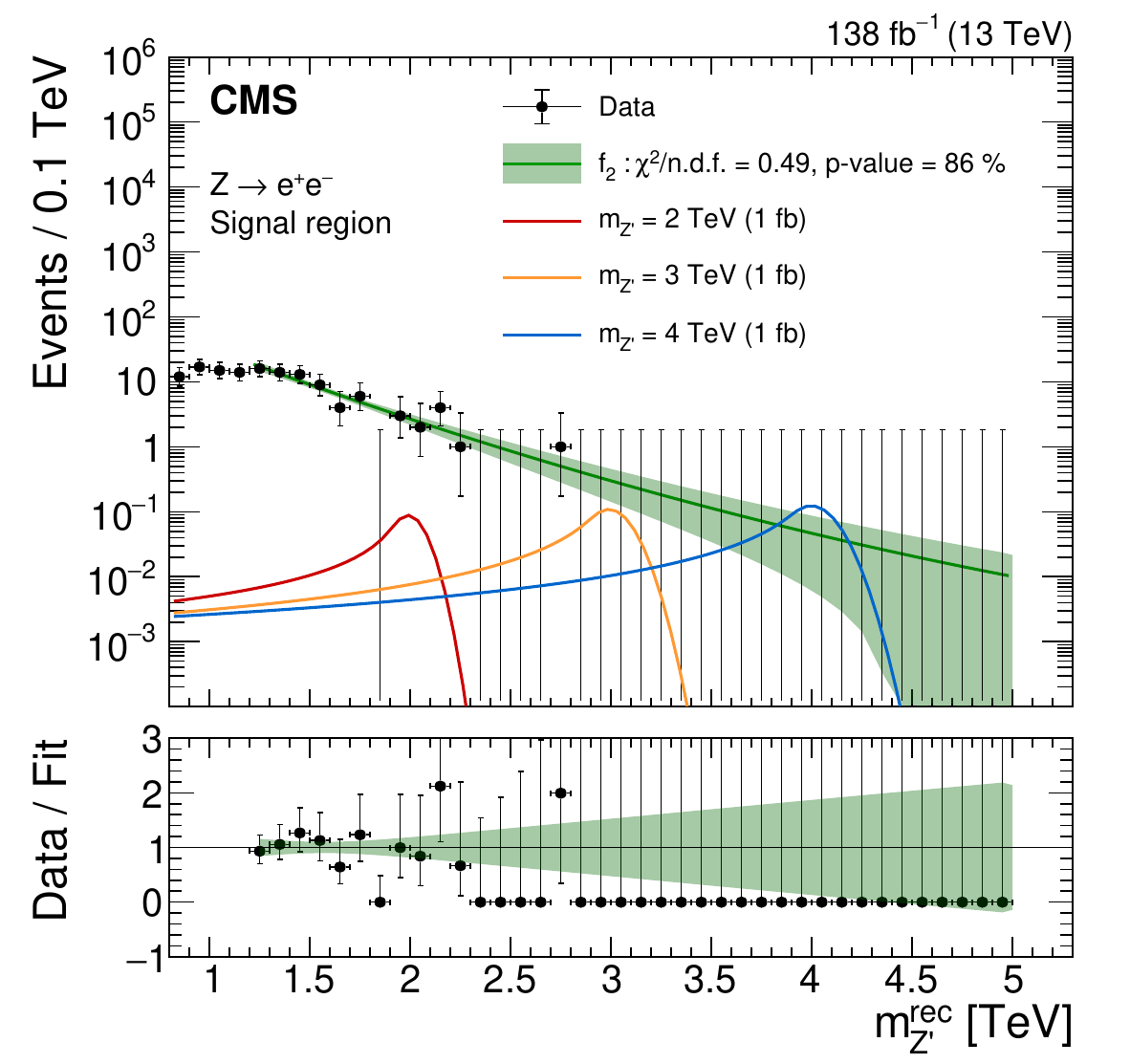}
\includegraphics[width=\cmsFigWidth]{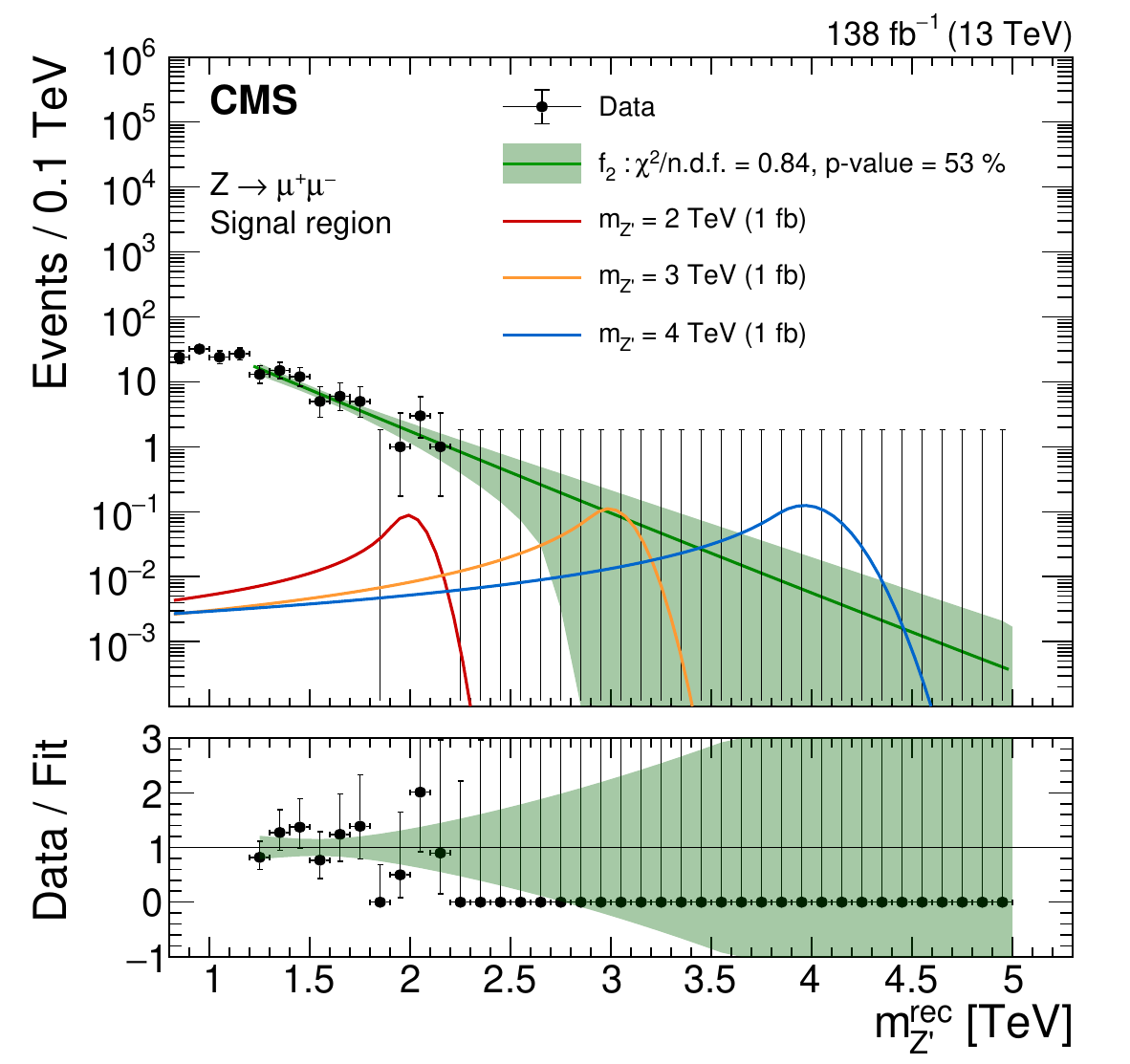}\\
\includegraphics[width=\cmsFigWidth]{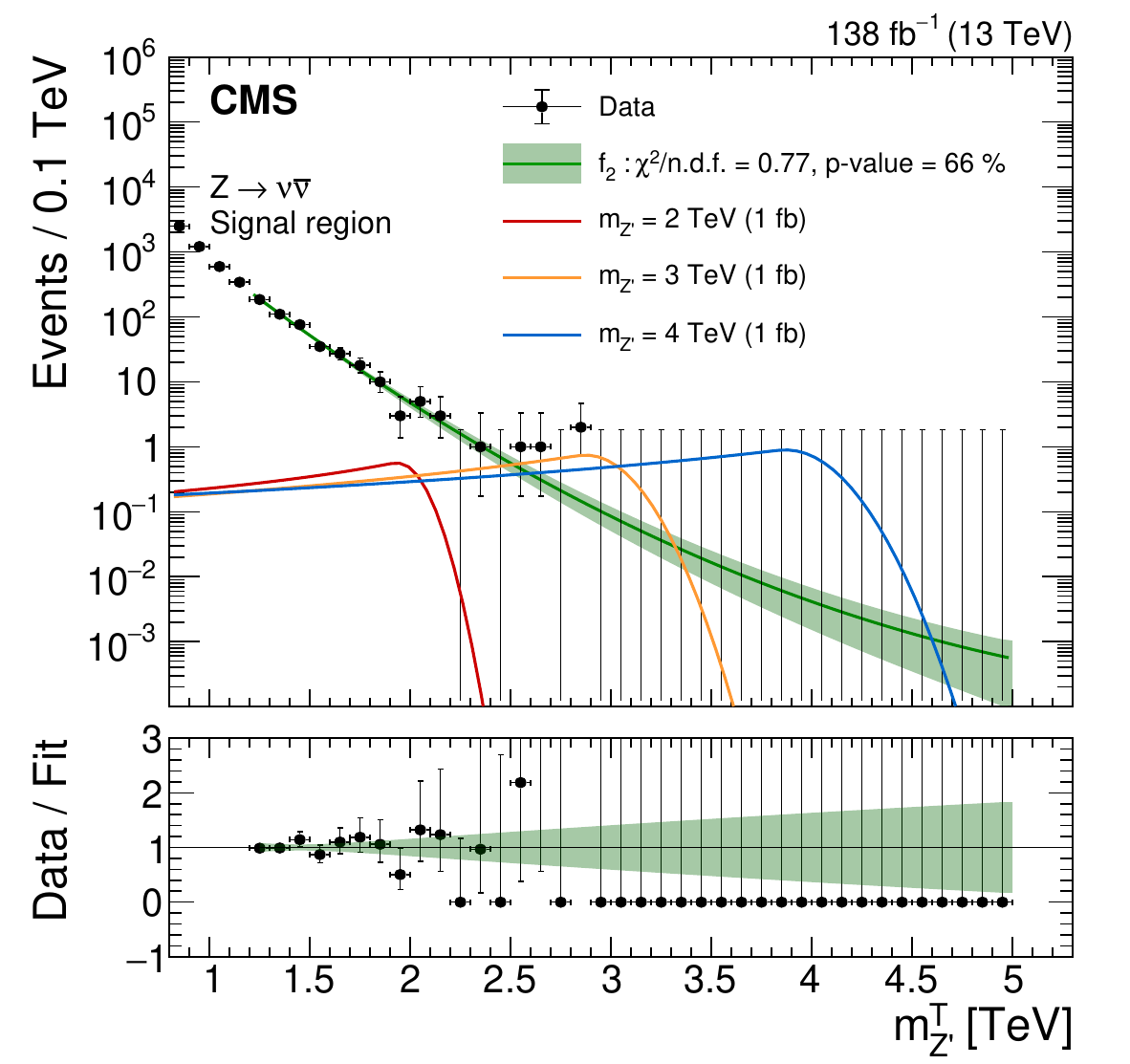}
\caption{Distributions in \mZprimeRec and \mTZprime for data in the SRs, together with 
fits of the background functions under the background-only hypothesis for the electron (upper left),
muon (upper right), and neutrino (lower) channels.
The signal predictions are shown for different \Zprime boson masses,
normalized to an arbitrary cross section of 1\unit{fb}.
In the panels below the distributions, the ratios of data to the background function are displayed.
The shaded green areas represent the statistical uncertainty from the fit.
The $\chi^{2}$ values per number of degrees of freedom ($\chi^{2}$/n.d.f.) and 
the corresponding $p$-values are provided for each fit.
\label{fig:ZprimeZH_DATASR}
}
\end{figure}

We use the \mZprime and \mTZprime distributions measured in the three SRs 
for the statistical interpretation.
The distributions in data are shown in Fig.~\ref{fig:ZprimeZH_DATASR}.
The background functions have been obtained from a fit with the background-only hypothesis, where 
the parameters of the fit functions have been left unconstrained to ensure an unbiased result.
The data are in agreement with the background-only fits.

\begin{figure}[p]
\centering
\includegraphics[width=0.75\textwidth]{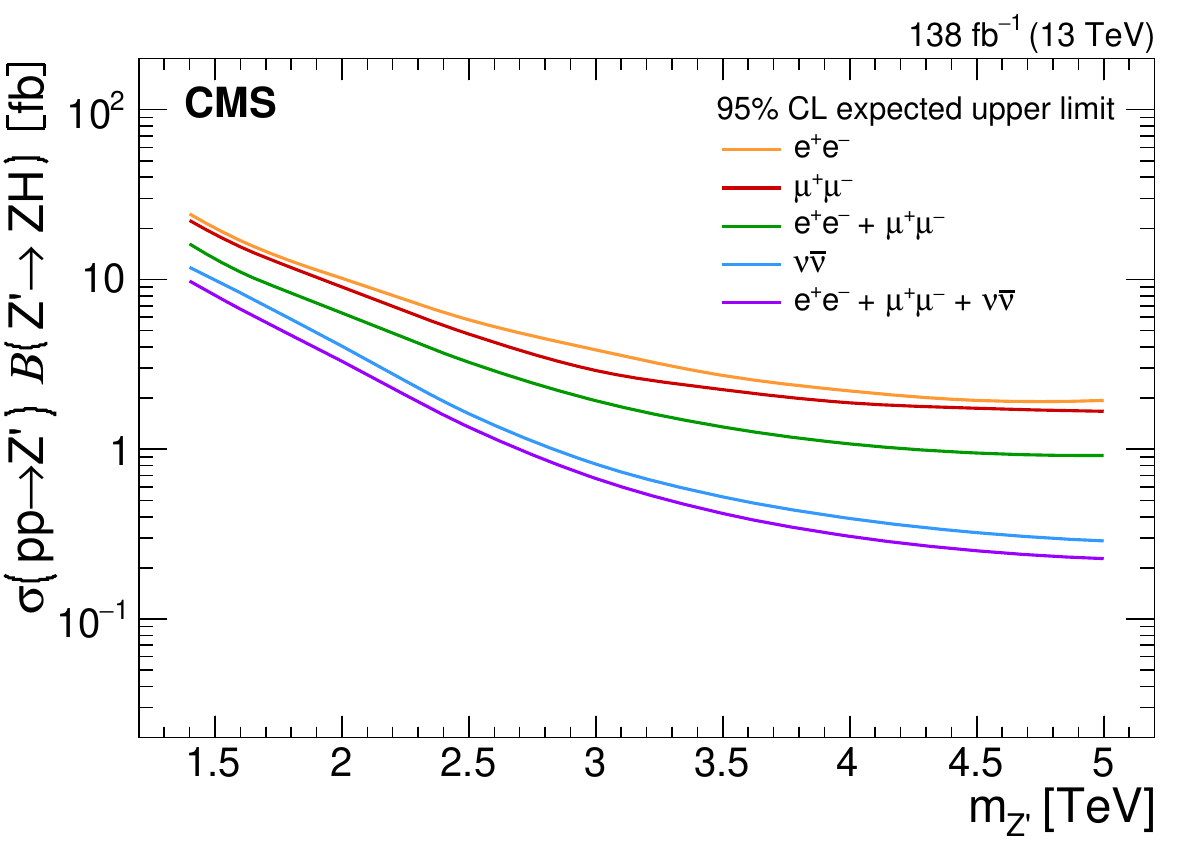}
\caption{Expected upper limits at 95\% \CL on the product of the production cross section
\Zprimexsec and branching fraction \ZprimeBR as functions of the \Zprime boson mass.
The lines correspond to the three different final states and their combinations.
\label{fig:ZprimeZH_Limits_channels}
}
\end{figure}
\begin{figure}[p]
\centering
\includegraphics[width=0.75\textwidth]{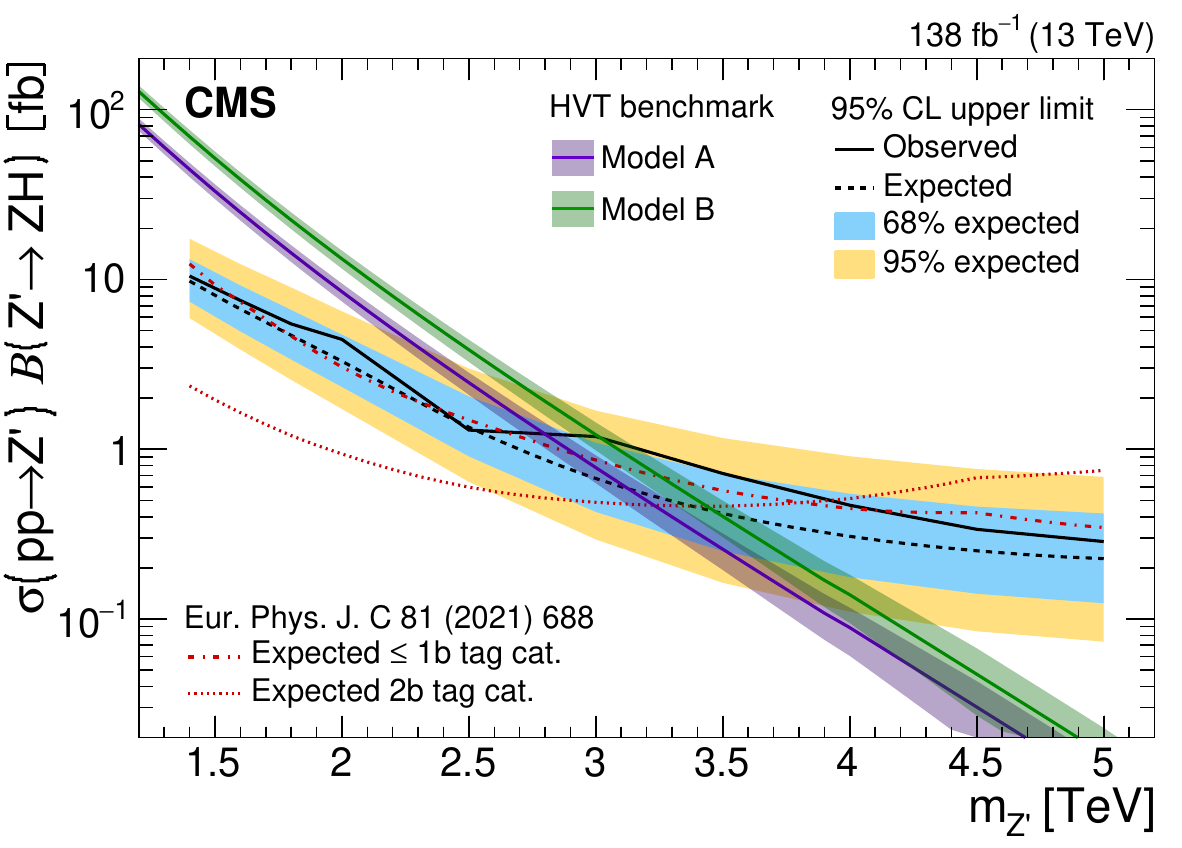}
\caption{Expected and observed upper limits at 95\% \CL on the product of the production cross section
\Zprimexsec and branching fraction \ZprimeBR as functions of the \Zprime boson mass from the combination of all final states.
The limits are compared with the predictions of the HVT model and the expected limits (shown by red curves) 
from a previous analysis~\cite{CMS:2021fyk}.
\label{fig:ZprimeZH_Limits}
}
\end{figure}
We set upper limits at 95\%~\CL on
the product $\Zprimexsec\ZprimeBR$ as functions of \mZprime.
The expected 95\% \CL limits obtained for the individual channels and their combination are shown in Fig.~\ref{fig:ZprimeZH_Limits_channels}.
The neutrino channel is the most sensitive channel over the entire mass range because of the larger branching fraction 
and the higher selection efficiency.

The final expected and observed exclusion limits at 95\% \CL
resulting from the combination of the muon, 
electron, and neutrino channels are shown in Fig.~\ref{fig:ZprimeZH_Limits}.
The observed 95\% \CL limits agree with the expectations from the background-only hypothesis within 
about one standard deviation over the full mass range considered.
We observe an improvement in the sensitivity as compared with
the non-VBF production category studied in the previous CMS analysis~\cite{CMS:2021fyk}.
The analysis has better expected sensitivity than the \onebcat from the previous 
analysis by up to about 60\% on $\Zprimexsec\ZprimeBR$. The analysis has better sensitivity than the
\twobcat for $\mZprime>3.4\TeV$, where previously this crossing point was at 3.85\TeV.
The lower \Zprime boson mass limits from this analysis alone are at 2.8 and 3\TeV
for Models~A and B, respectively.
We used a combination of \PQb, \PQc, and light-flavour tagging for hadronic \PH boson final states
to improve the sensitivity of a BSM search at high resonance masses.
This approach enhances \PH boson decays into \ccbar and $\PV\PV^{*}\to\text{4 quarks}$
while also recovering misidentified $\PH\to\bbbar$ decays.

\begin{figure}[tb]
\centering
\includegraphics[width=0.85\textwidth]{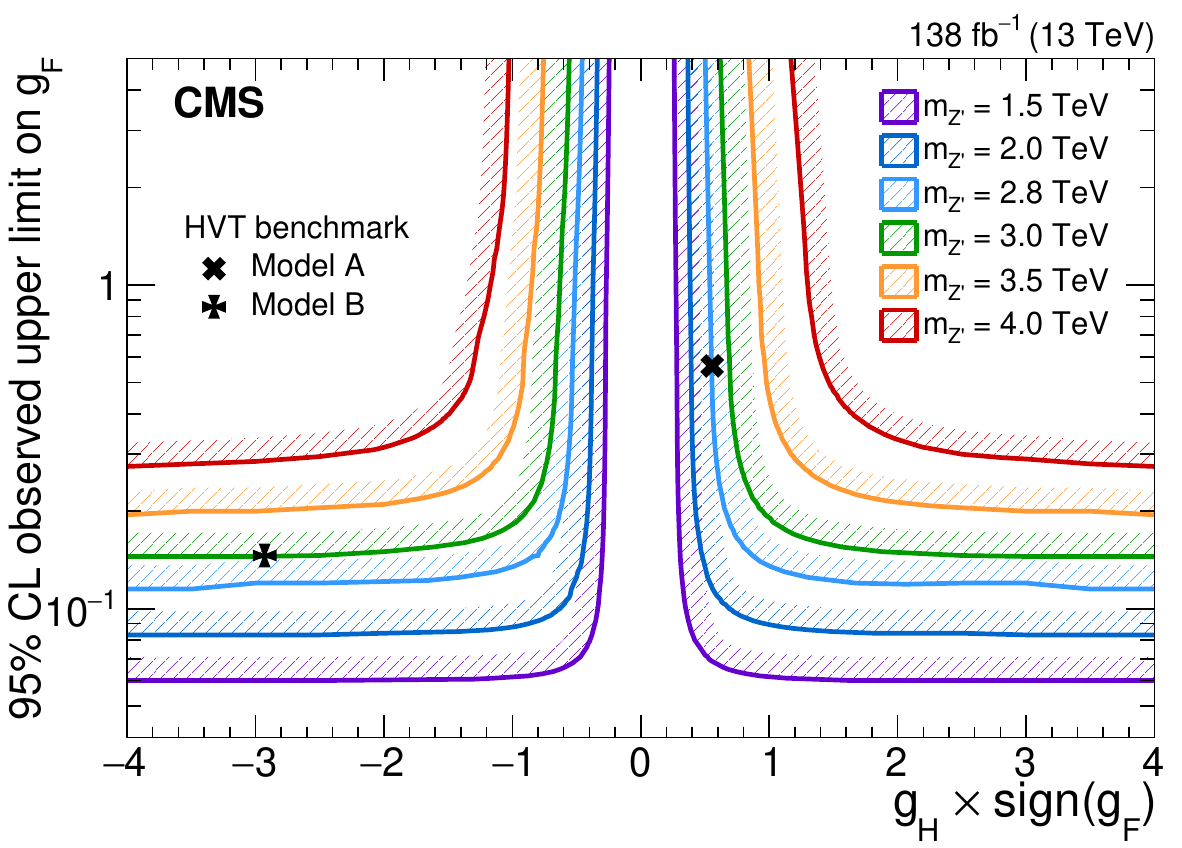}
\caption{Observed upper limits at 95\% \CL on \gF for different \Zprime boson masses as functions
of the product of \gH with the sign of \gF.
The two benchmark scenarios of the HVT model are shown by the black markers.
\label{fig:Limits_coupling}
}
\end{figure}

The upper limits on the cross sections are translated into two-dimensional upper limits 
on the coupling parameters for fermions (\gF), and bosons (\gH) in the HVT model.
The resulting exclusion contours are shown in Fig.~\ref{fig:Limits_coupling}.

\section{Summary} \label{sec:Summary}

A search has been presented for the production of a heavy resonance 
with mass in the range of 1.4--5\TeV that decays into a \PZ and a Higgs (\PH) boson.
The analysis is performed using data recorded
with the CMS detector at a centre-of-mass energy of 13\TeV,
corresponding to an integrated luminosity of 138\fbinv.
Exclusion limits at 95\% confidence level are set on both the mass 
of a heavy resonance and the couplings to fermions and bosons in the heavy vector triplet model.
Resonances with masses below 3\TeV are excluded.

The analysis focuses on \PZ boson decays into a pair of electrons, muons, or neutrinos,
and the hadronic decays of the \PH boson reconstructed as a single large-radius jet.
A novel approach analyzing the flavour content and substructure of the \PH boson jet 
is deployed to improve the sensitivity for high resonance masses. 
This analysis employs state-of-the-art algorithms to identify hadronic \PH boson decays. 
The use of these advanced taggers significantly improves the signal-to-background 
ratio for \PH boson jets with less than two \PQb-tagged subjets.  
The signal \PH boson jets comprise about 60, 20, and 15\% $\PH\to\bbbar$, $\qqbar\qqbar$, and \ccbar 
decays, respectively, where the $\PH\to\bbbar$ decays correspond to jets 
that are rejected by a selection based on two \PQb-tagged subjets and are recovered 
by this analysis. 
A significantly improved sensitivity for high resonance masses in the $\PZ\PH$ channel 
is provided by the new techniques employed by this analysis.

\begin{acknowledgments}
    We congratulate our colleagues in the CERN accelerator departments for the excellent performance of the LHC and thank the technical and administrative staffs at CERN and at other CMS institutes for their contributions to the success of the CMS effort. In addition, we gratefully acknowledge the computing centres and personnel of the Worldwide LHC Computing Grid and other centres for delivering so effectively the computing infrastructure essential to our analyses. Finally, we acknowledge the enduring support for the construction and operation of the LHC, the CMS detector, and the supporting computing infrastructure provided by the following funding agencies: SC (Armenia), BMBWF and FWF (Austria); FNRS and FWO (Belgium); CNPq, CAPES, FAPERJ, FAPERGS, and FAPESP (Brazil); MES and BNSF (Bulgaria); CERN; CAS, MoST, and NSFC (China); MINCIENCIAS (Colombia); MSES and CSF (Croatia); RIF (Cyprus); SENESCYT (Ecuador); ERC PRG, RVTT3 and MoER TK202 (Estonia); Academy of Finland, MEC, and HIP (Finland); CEA and CNRS/IN2P3 (France); SRNSF (Georgia); BMBF, DFG, and HGF (Germany); GSRI (Greece); NKFIH (Hungary); DAE and DST (India); IPM (Iran); SFI (Ireland); INFN (Italy); MSIP and NRF (Republic of Korea); MES (Latvia); LMTLT (Lithuania); MOE and UM (Malaysia); BUAP, CINVESTAV, CONACYT, LNS, SEP, and UASLP-FAI (Mexico); MOS (Montenegro); MBIE (New Zealand); PAEC (Pakistan); MES and NSC (Poland); FCT (Portugal); MESTD (Serbia); MCIN/AEI and PCTI (Spain); MOSTR (Sri Lanka); Swiss Funding Agencies (Switzerland); MST (Taipei); MHESI and NSTDA (Thailand); TUBITAK and TENMAK (Turkey); NASU (Ukraine); STFC (United Kingdom); DOE and NSF (USA).

    \hyphenation{Rachada-pisek} Individuals have received support from the Marie-Curie programme and the European Research Council and Horizon 2020 Grant, contract Nos.\ 675440, 724704, 752730, 758316, 765710, 824093, 101115353, 101002207, and COST Action CA16108 (European Union); the Leventis Foundation; the Alfred P.\ Sloan Foundation; the Alexander von Humboldt Foundation; the Science Committee, project no. 22rl-037 (Armenia); the Belgian Federal Science Policy Office; the Fonds pour la Formation \`a la Recherche dans l'Industrie et dans l'Agriculture (FRIA-Belgium); the F.R.S.-FNRS and FWO (Belgium) under the ``Excellence of Science -- EOS" -- be.h project n.\ 30820817; the Beijing Municipal Science \& Technology Commission, No. Z191100007219010 and Fundamental Research Funds for the Central Universities (China); the Ministry of Education, Youth and Sports (MEYS) of the Czech Republic; the Shota Rustaveli National Science Foundation, grant FR-22-985 (Georgia); the Deutsche Forschungsgemeinschaft (DFG), among others, under Germany's Excellence Strategy -- EXC 2121 ``Quantum Universe" -- 390833306, and under project number 400140256 - GRK2497; the Hellenic Foundation for Research and Innovation (HFRI), Project Number 2288 (Greece); the Hungarian Academy of Sciences, the New National Excellence Program - \'UNKP, the NKFIH research grants K 131991, K 133046, K 138136, K 143460, K 143477, K 146913, K 146914, K 147048, 2020-2.2.1-ED-2021-00181, and TKP2021-NKTA-64 (Hungary); the Council of Science and Industrial Research, India; ICSC -- National Research Centre for High Performance Computing, Big Data and Quantum Computing and FAIR -- Future Artificial Intelligence Research, funded by the NextGenerationEU program (Italy); the Latvian Council of Science; the Ministry of Education and Science, project no. 2022/WK/14, and the National Science Center, contracts Opus 2021/41/B/ST2/01369 and 2021/43/B/ST2/01552 (Poland); the Funda\c{c}\~ao para a Ci\^encia e a Tecnologia, grant CEECIND/01334/2018 (Portugal); the National Priorities Research Program by Qatar National Research Fund; MCIN/AEI/10.13039/501100011033, ERDF ``a way of making Europe", and the Programa Estatal de Fomento de la Investigaci{\'o}n Cient{\'i}fica y T{\'e}cnica de Excelencia Mar\'{\i}a de Maeztu, grant MDM-2017-0765 and Programa Severo Ochoa del Principado de Asturias (Spain); the Chulalongkorn Academic into Its 2nd Century Project Advancement Project, and the National Science, Research and Innovation Fund via the Program Management Unit for Human Resources \& Institutional Development, Research and Innovation, grant B39G670016 (Thailand); the Kavli Foundation; the Nvidia Corporation; the SuperMicro Corporation; the Welch Foundation, contract C-1845; and the Weston Havens Foundation (USA).
\end{acknowledgments}

\bibliography{auto_generated}\cleardoublepage \appendix\section{The CMS Collaboration \label{app:collab}}\begin{sloppypar}\hyphenpenalty=5000\widowpenalty=500\clubpenalty=5000
\cmsinstitute{Yerevan Physics Institute, Yerevan, Armenia}
{\tolerance=6000
A.~Hayrapetyan, A.~Tumasyan\cmsAuthorMark{1}\cmsorcid{0009-0000-0684-6742}
\par}
\cmsinstitute{Institut f\"{u}r Hochenergiephysik, Vienna, Austria}
{\tolerance=6000
W.~Adam\cmsorcid{0000-0001-9099-4341}, J.W.~Andrejkovic, T.~Bergauer\cmsorcid{0000-0002-5786-0293}, S.~Chatterjee\cmsorcid{0000-0003-2660-0349}, K.~Damanakis\cmsorcid{0000-0001-5389-2872}, M.~Dragicevic\cmsorcid{0000-0003-1967-6783}, P.S.~Hussain\cmsorcid{0000-0002-4825-5278}, M.~Jeitler\cmsAuthorMark{2}\cmsorcid{0000-0002-5141-9560}, N.~Krammer\cmsorcid{0000-0002-0548-0985}, A.~Li\cmsorcid{0000-0002-4547-116X}, D.~Liko\cmsorcid{0000-0002-3380-473X}, I.~Mikulec\cmsorcid{0000-0003-0385-2746}, J.~Schieck\cmsAuthorMark{2}\cmsorcid{0000-0002-1058-8093}, R.~Sch\"{o}fbeck\cmsorcid{0000-0002-2332-8784}, D.~Schwarz\cmsorcid{0000-0002-3821-7331}, M.~Sonawane\cmsorcid{0000-0003-0510-7010}, W.~Waltenberger\cmsorcid{0000-0002-6215-7228}, C.-E.~Wulz\cmsAuthorMark{2}\cmsorcid{0000-0001-9226-5812}
% ...
\par}
\vskip\cmsinstskip
\dag:~Deceased\\
$^{1}$Also at Yerevan State University, Yerevan, Armenia\\
$^{2}$Also at TU Wien, Vienna, Austria\\
% ...
\end{sloppypar}
%%% END EDITABLE REGION %%%
% skeleton_end
\end{document}